\documentclass[11pt]{article}
\usepackage{amssymb}
\makeatletter\@addtoreset {equation}{section}\makeatother

\newtheorem{theorem}{Theorem}
\newtheorem{algorithm}{Algorithm}
\newtheorem{example}{Example}

\usepackage[dvips]{epsfig}
\usepackage{graphicx}

\begin{document}

\title{\bf Lyapunov--Schmidt reduction algorithm for three-dimensional discrete vortices}
\author{Mike Lukas$^{\dagger}$, Dmitry Pelinovsky$^{\dagger}$ and P.G. Kevrekidis$^{\dagger  \dagger}$ \\
{\small $^{\dagger}$ Department of Mathematics, McMaster
University, Hamilton, Ontario, Canada, L8S 4K1} \\
{\small $^{\dagger \dagger}$ Department of Mathematics and
Statistics, University of Massachusetts, Amherst, MA 01003 } }
\date{\today}
\maketitle

\begin{abstract}
We address persistence and stability of three-dimensional vortex
configurations in the discrete nonlinear Schr\"{o}dinger (NLS)
equation and develop a symbolic package based on Wolfram's
MATHEMATICA for computations of the Lyapunov--Schmidt reduction
method. The Lyapunov--Schmidt reduction method is a theoretical
tool which enables us to study continuations and terminations of
the discrete vortices for small coupling between lattice nodes as
well as the spectral stability of the persistent configurations.
The method was developed earlier in the context of the
two-dimensional NLS lattice and applied to the on-site and
off-site configurations (called the vortex cross and the vortex
cell) by using semi-analytical computations \cite{PKF05,KP06}. The
present treatment develops a full symbolic computational package
which takes a desired waveform at the anti-continuum limit of
uncoupled sites, performs a required number of Lyapunov--Schmidt
reductions and outputs the predictions on whether the
configuration persists, for finite coupling, in the
three-dimensional lattice and whether it is stable or unstable. It
also provides approximations for the eigenvalues of the linearized
stability problem. We report a number of applications of the
algorithm to important multi-site configurations, such as the
simple cube, the double cross, and the diamond. For each
three-dimensional configuration, we identify exactly one solution,
which is stable for small coupling between lattice nodes.
\end{abstract}

\section{Introduction}

Over the last decade, the topic of nonlinear dynamical lattices
and of the coherent structures that arise in them has attracted
considerable attention. This can be mainly attributed
to a diverse set of research themes where the corresponding
mathematical models have emerged as an appropriate description
of the physical problem. Such areas include, but are not limited
to, arrays of nonlinear-optical waveguides \cite{optics} and photorefractive
crystal lattices \cite{moti}, Bose-Einstein condensates (BECs)
trapped in optical lattices (OLs) \cite{bec_reviews},
Josephson-junction ladders \cite{alex}, micro-mechanical models of
cantilever arrays \cite{sievers} or even simple models of the
complex dynamics of the DNA double strand \cite{peyrard}.
These, in turn, have spurted an increasing mathematical interest
in Hamiltonian discrete systems and produced a considerable volume
of theoretical results, which has partially been summarized
in the reviews \cite{reviews}.

Arguably, the most prototypical among the discrete nonlinear
Hamiltonian models is the discrete nonlinear Schr{\"o}dinger
equation (DNLS) \cite{dnls}, which results from a
centered-difference discretization of its famous continuum analog,
the nonlinear Schr{\"o}dinger equation (NLS) \cite{sulem}. The
DNLS arises most notably in the nonlinear optics of waveguide
arrays, where it was developed \cite{christo} as an envelope wave
description of the electric field in each of the waveguides. It
was also systematically derived as a tight-binding model for
Bose-Einstein condensates trapped in, so-called, optical lattices
\cite{alfimov}. In other settings, such as e.g. photorefractive
crystals, the validity of the DNLS model is more limited, yet
still a lot of its qualitative features can be observed even
experimentally. It is for this reason that this simple yet rich
model has been used to justify numerous experimental observations
in the above-mentioned areas, including the formation of discrete
solitons \cite{eis}, the existence of Peierls-Nabarro barriers for
such waves \cite{PN}, the modulational instability of uniform
states both in optical \cite{dnc0} and in BEC \cite{as}
experiments, and the formation of discrete vortices \cite{dv} in
two-dimensions, among many others. In the earlier work of
\cite{PKF1} and \cite{PKF05,KP06}, we studied systematically the
solutions that can be obtained in the one- and two-dimensional
installments of the model respectively, using a combination of
methods of Lyapunov-Schmidt reductions and perturbation theory in
the, so-called, anti-continuum limit of lattice sites uncoupled
with each other.

In the present paper, we provide a mathematical framework and a
symbolic package for the systematic computations of the existence
and the stability of localized states in the fully
three-dimensional DNLS model. To the best of our knowledge, this
is the first time that systematic stability results are developed
for the three-dimensional system. Earlier numerical work in
\cite{pgk1,pgk2} revealed a considerable wealth of possible
states, including octupoles, diamonds and vortex cubes, among many
others. It is these states and different variants thereof that our
systematic methodology addresses in the present publication, by
formulating a symbolic approach that can be straightforwardly used
to study the potential persistence and linear stability of {\it
any} desired configuration.

Our results on the three-dimensional DNLS model cannot be applied
to dynamics of waveguide arrays since the third spatial direction
in the optical problem represents the evolution time variable.
However, the model is still relevant to BECs and an additional
possibility for its physical realization is offered by a
three-dimensional crystal built of microresonators \cite{photons}.

Our article is structured as follows. Section 2 formulates the
mathematical problem of interest. Section 3 presents the main
results for the development of the computational algorithm based on
the Lyapunov--Schmidt reduction method. Section 4 reports the
application of the computational algorithm to three kinds of
three-dimensional vortex configurations. Section 5 compares these
results with direct numerical bifurcation results of the three-dimensional
DNLS model. Section 6 concludes the paper. Appendix A contains
typical outputs of the MATHEMATICA software package.

\section{Formulation of the problem}
\label{section-formalism}

We consider the discrete nonlinear Schr\"{o}dinger (NLS) equation
in three spatial dimensions:
\begin{equation}
\label{3NLS} i \dot{u}_n + \epsilon \Delta u_n + |u_n|^2 u_n = 0,
\qquad n \in \mathbb{Z}^3, \;\; t \in \mathbb{R}_+, \;\; u_n \in
\mathbb{C},
\end{equation}
where $\epsilon > 0$ is the coupling strength (which can also be
thought of as the reciprocal squared lattice spacing) and $\Delta
u_n$ is the discrete three-dimensional Laplacian
$$
\Delta u_n = u_{n+e_1} + u_{n-e_1} + u_{n+e_2} + u_{n-e_2} +
u_{n+e_3} + u_{n-e_3} - 6 u_{n}, \qquad n \in \mathbb{Z}^3,
$$
with $\{ e_1,e_2,e_3\}$ being standard unit vectors on
$\mathbb{Z}^3$.  The discrete NLS equation (\ref{3NLS}) is a
Hamiltonian system with the Hamiltonian function
\begin{equation}
\label{Hamiltonian-system} H = \epsilon \sum_{j=1}^3 \| u_{n+e_j}
- u_n \|^2_{l^2(\mathbb{Z}^3)} - \frac{1}{2} \| u_n
\|^4_{l^4(\mathbb{Z}^3)},
\end{equation}
which is referred to as the energy of the discrete system. Due to
invariance of the discrete NLS equation (\ref{3NLS}) with respect
to time translation $u_n(t) \mapsto u_n(t-t_0)$, $\forall t_0 \in
\mathbb{R}$, the energy $H$ is constant in time, e.g. $H(t) =
H(0)$. Another conserved quantity is the discrete $l^2$-norm, such
that $Q = \| u_n(t) \|^2_{l^2(\mathbb{Z}^3)}$ and $Q(t) = Q(0)$,
which is related to the invariance of the NLS equation
(\ref{3NLS}) with respect to the gauge transformation $u_{n}(t)
\mapsto u_{n}(t) e^{i \theta_0}$, $\forall \theta_0 \in
\mathbb{R}$. The natural phase space of the discrete system is $X
= l^2(\mathbb{Z}^3) \cap l^4(\mathbb{Z}^3)$.

Let ${\bf u}$ be an abstract vector for the triple-indexed
doubly-infinite two-component sequences of real and imaginary parts
of $\{ u_n \}_{n \in \mathbb{Z}^3}$, such that the $2$-block at the
node $n \in \mathbb{Z}^3$ is
$$
{\bf u}_n = \left[ \begin{array}{c} {\rm Re}(u_n) \\ {\rm Im}(u_n)
\end{array} \right].
$$
Then, the discrete NLS equation (\ref{3NLS}) is equivalent to the
dynamical system $\frac{d {\bf u}}{dt}  = J \nabla H[{\bf u}]$,
where $H[{\bf u}] : \; X \mapsto \mathbb{R}^1$ is the Hamiltonian
function (\ref{Hamiltonian-system}) and $(J,\nabla)$ are the
standard symplectic and gradient operators, respectively. Operators
$(J,\nabla)$ are block-diagonal with the $2$-block at the node $n
\in \mathbb{Z}^3$ being
$$
J_n = \left[ \begin{array}{cc} 0 & 1 \\ -1 & 0 \end{array}
\right], \qquad \nabla_n = \left[ \begin{array}{cc} \partial_{{\rm Re}(u_n)} \\
\partial_{{\rm Im}(u_n)} \end{array} \right]
$$

We consider the time-periodic (stationary) solutions of the
discrete NLS equation (\ref{3NLS}) in the form
\begin{equation}
\label{vortex} u_n(t) = \phi_n e^{i (1 - 6 \epsilon) t}, \qquad n
\in \mathbb{Z}^3, \;\; \phi_n \in \mathbb{C},
\end{equation}
where all parameters have been normalized for convenience. The
sequence $\{ \phi_n \}_{n \in \mathbb{Z}^3}$ is a solution of the
difference equation
\begin{equation}
\label{3difference} (1 - |\phi_n|^2) \phi_n = \epsilon \Sigma
\phi_n, \qquad \Sigma = \Delta + 6,   \qquad   n \in \mathbb{Z}^3,
\end{equation}
which is obtained by variation of the Lyapunov functional
$\Lambda[{\bf u}] = H[{\bf u}] + Q[{\bf u}]$, such that the
Euler--Lagrange equation $\nabla \Lambda[{\bf u}] |_{{\bf u} =
\mbox{\boldmath $\phi$}} = 0$ is equivalent to the difference
equation (\ref{3difference}).

If the localized solution $\mbox{\boldmath $\phi$}$ of the
difference equation (\ref{3difference}) exists, the time-evolution
of the discrete NLS lattice (\ref{3NLS}) near the time-periodic
space-localized solution (\ref{vortex}) is defined by the
linearization
\begin{equation}
\label{linearization} u_n(t) = e^{i (1 - 6 \epsilon) t} \left(
\phi_n + a_n e^{\lambda t} + \bar{b}_n e^{\bar{\lambda} t}
\right), \qquad n \in {\mathbb Z}^3,
\end{equation}
where $\lambda$ is the spectral parameter and the sequence $\{
(a_n,b_n)\}_{n \in \mathbb{Z}^3}$ solves the linear eigenvalue
problem for difference operators
\begin{eqnarray}
\label{linear-NLS} \left(1 - 2 |\phi_n|^2\right) a_n - \phi_n^2
b_n - \epsilon \Sigma a_n = i \lambda a_n, \;\; - \bar{\phi}_n^2
a_n + \left(1 - 2 |\phi_n|^2\right) b_n - \epsilon \Sigma b_n = -i
\lambda b_n, \quad n \in \mathbb{Z}^3.
\end{eqnarray}
If there exists a solution with ${\rm Re}(\lambda) > 0$, the
stationary solution (\ref{vortex}) is called {\em spectrally
unstable}. Otherwise, it is neutrally spectrally stable.

The difference equation (\ref{3difference}) with $\epsilon = 0$
has a general set of localized modes on $n \in \mathbb{Z}^3$:
\begin{equation}
\label{vortex-limit} \phi_n^{(0)} =
\left\{ \begin{array}{cc} e^{i \theta_n}, \quad n \in S, \\
0, \qquad n \in S^{\perp}, \end{array} \right.
\end{equation}
where $S$ is a bounded set of nodes on the lattice $n \in
\mathbb{Z}^3$, $S^{\perp} = \mathbb{Z}^3 \backslash S$, and $\{
\theta_n \}_{n \in S}$ is a set of phase configurations. The set
$\{ \theta_n \}_{n \in S}$ is arbitrary for $\epsilon = 0$. It is
called a {\em vortex configuration} if $\theta_{n_0} = 0$ for a
single node $n_0 \in S \subset \mathbb{Z}^3$ and $\theta_{n_1}
\neq \{0,\pi\}$ for at least one $n_1 \in S$, $n_1 \neq n_0$. If
$\theta_n \in \{ 0, \pi \}$, $\forall n \in S$, it is dubbed a
{\em discrete soliton}. We address here the main questions: {\em
For what vortex configurations, the localized mode
(\ref{vortex-limit}) can be continued in the difference equation
(\ref{3difference}) for small $\epsilon > 0$ and is stable in the
linearized discrete NLS equation (\ref{linear-NLS})?}

Although the questions and the algorithms of their solution are
rather general, we would like to consider specific vortex
configurations in the three-dimensional NLS lattice. Some of these
configurations have been addressed in the previous works
\cite{pgk1,pgk2} as they are thought to be elementary building
blocks of three-dimensional discrete structures. The three specific
configurations are formulated in terms of the limiting solution
(\ref{vortex-limit}).

\begin{itemize}
\item[(i)] {\em A simple cube} consists of two adjacent planes of aligned vortex
cells. More precisely, we set $S = S_0 \oplus S_1$, where
\begin{equation}
\label{contour1} S_l = \{ (0,0,l), (1,0,l), (1,1,l), (0,1,l) \},
\qquad l = 0,1,
\end{equation}
such that $N = {\rm dim}(S) = 8$. Let $j$ be the index enumerating
nodes in the contours $S_0$ and $S_1$ according to the order
(\ref{contour1}). Let $\theta_{l,j}$ be the corresponding phase in
the set $\{ \theta_n \}_{n \in S}$, where $l = 0,1$ and $j =
1,2,3,4$. Then, the simple cube vortex configurations are
\begin{equation}
\label{vortex1} \theta_{0,j} = \frac{\pi (j-1)}{2}, \qquad
\theta_{1,j} = \theta_0 + s_0 \frac{\pi (j-1)}{2}, \qquad j =
1,2,3,4,
\end{equation}
where $\theta_0 = \{ 0, \frac{\pi}{2}, \pi, \frac{3 \pi}{2} \}$
and $s_0 = \{ +1, -1\}$. Two configurations with $s_0 = -1$ are
redundant: the case $\theta_0 = \pi$ can be obtained from the case
$\theta_0 = 0$ by multiplication of $u_n$ by $i$ and so does the
case $\theta_0 = \frac{3\pi}{2}$ from the case $\theta_0 =
\frac{\pi}{2}$. In what follows, we only consider the six
irreducible vortex configurations and show that only three
configurations persist for $\epsilon \neq 0$ and only one
configuration with $\theta_0 = \pi$ and $s_0 = 1$ is linearly
stable.

\item[(ii)] {\em A double cross} consists of two symmetric planes of
aligned vortex crosses separated by an empty plane. More precisely,
we set $S = S_{-1} \oplus S_1$, where
\begin{equation}
\label{contour3} S_l = \{ (-1,0,l), (0,-1,l), (1,0,l), (0,1,l) \},
\qquad l = -1,1,
\end{equation}
such that $N = {\rm dim}(S) = 8$. By using the same convention as in
(i), the double cross vortex configurations are expressed by
\begin{equation}
\label{vortex3} \theta_{-1,j} = \frac{\pi (j-1)}{2}, \qquad
\theta_{1,j} = \theta_0 + s_0 \frac{\pi (j-1)}{2}, \qquad j =
1,2,3,4,
\end{equation}
where $\theta_0 = \{ 0, \frac{\pi}{2}, \pi, \frac{3 \pi}{2} \}$
and $s_0 = \{ +1, -1\}$. Similarly to the simple cube vortex
configurations, we will consider six irreducible vortex
configurations and show that three configurations persist for
$\epsilon \neq 0$ and only one configuration with $\theta_0 = \pi$
and $s_0 = 1$ is linearly stable.

\item[(iii)] {\em A diamond} consists of a quadrupole in a central plane,
surrounded by two symmetric central off-plane nodes. More precisely, we set
$S = S_{-1} \oplus S_0 \oplus S_1$, where
\begin{equation}
\label{contour2} S_0 = \{ (-1,0,0), (0,-1,0), (1,0,0), (0,1,0) \},
\qquad S_{\pm 1} = \{ (0,0,\pm 1) \},
\end{equation}
such that $N = {\rm dim}(S) = 6$. By using the same convention as
in (i), the diamond vortex configurations are expressed by
\begin{equation}
\label{vortex2} \theta_{0,j} = \pi (j-1), \qquad j = 1,2,3,4,
\qquad \theta_{\pm 1,0} = \theta_0^{\pm},
\end{equation}
where $\theta_0^{\pm} = \{ 0, \frac{\pi}{2}, \pi, \frac{3 \pi}{2}
\}$. Six configurations with $\theta_0^- >  \theta_0^+$ are
redundant as they can be obtained from the corresponding
configurations with $\theta_0^- < \theta_0^+$ by reflection:
$\theta_0^{\pm} \mapsto \theta_0^{\mp}$. Three other
configurations with $\theta_0^- = \theta_0^+ = \{ \pi,
\frac{3\pi}{2} \}$ and $\theta_0^- = \pi$, $\theta_0^+ =
\frac{3\pi}{2}$ can be obtained from the configurations
$\theta_0^- = \theta_0^+ = \{ 0, \frac{\pi}{2} \}$ and $\theta_0^-
= 0$, $\theta_0^+ = \frac{\pi}{2}$ by multiplication of $u_n$ by
$-1$. One more configuration with $\theta_0^- = 0$ and $\theta_0^+
= \frac{3\pi}{2}$ can be obtained from the configuration with
$\theta_0^- = 0$ and $\theta_0^+ = \frac{\pi}{2}$ by complex
conjugation. In what follows, we only consider the six irreducible
vortex configurations and show that only three configurations
persist for $\epsilon \neq 0$ and only one configurations with
$\theta_0^- = \frac{\pi}{2}$ and $\theta_0^+ = \frac{3 \pi}{2}$ is
stable.
\end{itemize}

\section{Review of the Lyapunov--Schmidt reduction algorithm}
\label{section-algorithm}

We review the main results of \cite{PKF05}, where the
Lyapunov--Schmidt reduction method is developed to answer the
questions outlined in Section \ref{section-formalism}.

Let ${\cal O}(0)$ be a small neighborhood of $\epsilon = 0$ on
$\mathbb{R}^1$. Let $N = {\rm dim}(S)$ and ${\cal T}$ be the torus
on $[0,2\pi]^N$ for the vector $\mbox{\boldmath $\theta$}$ of phase
components $\{ \theta_n \}_{n \in S}$. We define the nonlinear
vector field ${\bf F}(\mbox{\boldmath $\phi$},\epsilon)$ on
$\mbox{\boldmath $\phi$} \in X$ and $\epsilon \in \mathbb{R}^1$,
such that the $2$-block at the node $n \in \mathbb{Z}$ is written by
\begin{equation}
\label{nonlinear-vector-field} {\bf F}_n(\mbox{\boldmath
$\phi$},\epsilon) = \left[
\begin{array}{c} (1 - |\phi_n|^2) \phi_n - \epsilon \Sigma
\phi_n \\ (1 - |\phi_n|^2) \bar{\phi}_n - \epsilon \Sigma
\bar{\phi}_n  \end{array} \right].
\end{equation}
The Jacobian $D_{\mbox{\boldmath $\phi$}} {\bf F}(\mbox{\boldmath
$\phi$},\epsilon)$ of the nonlinear vector field ${\bf
F}(\mbox{\boldmath $\phi$},\epsilon)$ at the solution
$\mbox{\boldmath $\phi$}$ for each $\epsilon \in {\cal O}(0)$
coincides with the {\em linearized energy operator} ${\cal H}$
related to the quadratic form for the Lyapunov function
$\Lambda[{\bf u}] = H[{\bf u}] + Q[{\bf u}]$ such that
$$
\Lambda[{\bf u}] = \Lambda[\mbox{\boldmath $\phi$}] + \frac{1}{2}
(\mbox{\boldmath $\psi$}, {\cal H} \mbox{\boldmath $\psi$}) + {\rm
O}(\| \mbox{\boldmath $\psi$} \|^3_{X \times X}),
$$
where ${\bf u}$ is expressed by (\ref{linearization}) and the
$2$-block of $\mbox{\boldmath $\psi$} \in X \times X$ is defined at
the node $n \in \mathbb{Z}^3$ by
$$
\mbox{\boldmath $\psi$}_n = \left[ \begin{array}{cc} a_n \\ b_n
\end{array} \right].
$$
The matrix operator ${\cal H}$ on $\mbox{\boldmath $\psi$} \in X
\times X$ is not block-diagonal due to the presence of the shift
operator $\Sigma$. We can still use a formal notation ${\cal H}_n$
for the ``$2$-block" of ${\cal H}$ at the node $n \in \mathbb{Z}^3$
\begin{equation}
\label{energy} {\cal H}_{n} = \left( \begin{array}{cc} 1 - 2
|\phi_n|^2 & - \phi_n^2 \\ - \bar{\phi}_n^2 & 1 - 2 |\phi_n|^2
\end{array} \right) - \epsilon \left( s_{+e_1} + s_{-e_1} +
s_{+e_2} + s_{-e_2} + s_{+e_3} + s_{-e_3} \right) \left(
\begin{array}{cc} 1 & 0 \\ 0 & 1 \end{array} \right),
\end{equation}
where $s_{e_j} u_n = u_{n+e_j}$ for $\{ e_1,e_2,e_3\} \in
\mathbb{Z}^3$. This notation allows us to write the matrix-vector
form ${\cal H} \mbox{\boldmath $\psi$}$ in the component form ${\cal
H}_n \mbox{\boldmath $\psi$}_n$ at each node $n \in \mathbb{Z}^3$.
In particular, the linear eigenvalue problem (\ref{linear-NLS}) can
be rewritten in the matrix-vector form
\begin{equation}
\label{eigenvalue} \sigma {\cal H} \mbox{\boldmath $\psi$} = i
\lambda \mbox{\boldmath $\psi$},
\end{equation}
where the $2$-block of $\sigma$ is a diagonal matrix of $(1,-1)$ at
each node $n \in \mathbb{Z}^3$.

Let $\mbox{\boldmath $\phi$}^{(0)} = \mbox{\boldmath
$\phi$}^{(0)}(\mbox{\boldmath $\theta$})$ be the solution
(\ref{vortex-limit}). By explicit computations, we have ${\bf
F}(\mbox{\boldmath $\phi$}^{(0)},0) = {\bf 0}$ and ${\rm Ker}({\cal
H}^{(0)}) = {\rm Span}(\{ {\bf e}_n\}_{n \in S}) \subset X \times
X$, where ${\cal H}^{(0)} = D_{\mbox{\boldmath $\phi$}} {\bf
F}(\mbox{\boldmath $\phi$}^{(0)},0)$ is block-diagonal with the
$2$-block at the node $n \in \mathbb{Z}^3$ given by
$$
({\cal H}^{(0)})_n = \left[ \begin{array}{cc} 1 & 0 \\ 0 & 1
\end{array} \right], \; n \in S^{\perp}, \qquad
({\cal H}^{(0)})_n = \left[ \begin{array}{cc} -1 & -e^{2 i \theta_n} \\
-e^{-2 i \theta_n} & -1 \end{array} \right], \; n \in S.
$$
We note that ${\cal H}^{(0)} {\bf e}_n = {\bf 0}$ and ${\cal
H}^{(0)} \hat{\bf e}_n = -2 \hat{\bf e}_n$, $n \in S \subset
\mathbb{Z}^3$, where the $2$-blocks of eigenvectors  ${\bf e}_n$
and $\hat{\bf e}_n$ at the node $k \in \mathbb{Z}^3$ are given by
$$
({\bf e}_n)_k = i \left[ \begin{array}{c} e^{i \theta_n} \\ -
e^{-i \theta_n}\end{array} \right] \; \delta_{k,n}, \qquad
(\hat{\bf e}_n)_k = \left[ \begin{array}{c} e^{i \theta_n} \\
e^{-i \theta_n}\end{array} \right] \; \delta_{k,n},
$$
with $\delta_{k,n}$ being a standard Kroneker symbol. Let ${\cal P}
: X \times X \mapsto {\rm Ker}({\cal H}^{(0)})$ be an orthogonal
projection operator to the $N$-dimensional kernel of ${\cal
H}^{(0)}$. In the explicit form, the projection operator ${\cal P}$
is expressed by
\begin{equation}
\label{projection-H-0} \forall {\bf f} = ({\bf f}_1,{\bf f}_2) \in X
\times X : \qquad ({\cal P} {\bf f})_n = \frac{\left( {\bf e}_n,
{\bf f} \right)}{\left( {\bf e}_n, {\bf e}_n \right)} = \frac{1}{2i}
\left( e^{-i \theta_n} ({\bf f}_1)_n - e^{i \theta_n}({\bf f}_2)_n
\right), \qquad n \in S.
\end{equation}
We note that the constraint ${\bf f}_2 = \bar{\bf f}_1$ is added
when solutions of the nonlinear vector equation ${\bf
F}(\mbox{\boldmath $\phi$},\epsilon) = {\bf 0}$ are considered
with $\mbox{\boldmath $\phi$} \in X$ and $\bar{\mbox{\boldmath
$\phi$}} \in X$.

Since the operator ${\cal H}^{(0)}$ is a self-adjoint Fredholm
operator of zero index, the decomposition $X \times X = {\rm
Ker}({\cal H}^{(0)}) \oplus {\rm Range}({\cal H}^{(0)})$ is
well-defined and so is the projection operator $({\cal I} - {\cal
P}) : X \times X \mapsto {\rm Range}({\cal H}^{(0)})$. By using the
Lyapunov--Schmidt reduction algorithm, we consider the decomposition
\begin{eqnarray}
\mbox{\boldmath $\phi$} = \mbox{\boldmath
$\phi$}^{(0)}(\mbox{\boldmath $\theta$}) + \sum_{n \in S} \alpha_n
{\bf e}_n + \mbox{\boldmath $\varphi$} \in X,
\end{eqnarray}
where $(\mbox{\boldmath $\varphi$},\bar{\mbox{\boldmath $\varphi$}})
\in {\rm Range}({\cal H}^{(0)})$ and $\alpha_n \in \mathbb{R}$ for
each $n \in S$. We note that
\begin{equation}
\label{derivative-0} \forall \mbox{\boldmath $\theta$}_0 \in {\cal
T} : \quad \mbox{\boldmath $\phi$}^{(0)}(\mbox{\boldmath
$\theta$}_0) + \sum_{n \in S} \alpha_n {\bf e}_n = \mbox{\boldmath
$\phi$}^{(0)}(\mbox{\boldmath $\theta$}_0 + \mbox{\boldmath
$\alpha$}) + {\rm O}(\|\mbox{\boldmath
$\alpha$}\|^2_{\mathbb{R}^n}).
\end{equation}
Since the values of $\mbox{\boldmath $\theta$}$ in
$\mbox{\boldmath $\phi$}^{(0)}(\mbox{\boldmath $\theta$})$ have
not been defined yet, we can set $\alpha_n = 0$, $\forall n \in S$
without loss of generality. The splitting equations in the
Lyapunov--Schmidt reduction algorithm are
$$
{\cal P} {\bf F}(\mbox{\boldmath $\phi$}^{(0)}(\mbox{\boldmath
$\theta$})+\mbox{\boldmath $\varphi$},\epsilon) = 0, \qquad ({\cal
I} - {\cal P}) {\bf F}(\mbox{\boldmath
$\phi$}^{(0)}(\mbox{\boldmath $\theta$})+\mbox{\boldmath
$\varphi$},\epsilon) = 0.
$$
We note that $({\cal I} - {\cal P}) {\cal H} ({\cal I} - {\cal P}) :
{\rm Range}({\cal H}^{(0)}) \mapsto {\rm Range}({\cal H}^{(0)})$ is
analytic in $\epsilon \in {\cal O}(0)$ and invertible at $\epsilon =
0$, while ${\bf F}(\mbox{\boldmath $\phi$},\epsilon)$ is analytic in
$\epsilon \in {\cal O}(0)$. By the Implicit Function Theorem for
Analytic Vector Fields, there exists a unique solution
$\mbox{\boldmath $\varphi$} \in X$ analytic in $\epsilon \in {\cal
O}(0)$ and dependent on $\mbox{\boldmath $\theta$} \in {\cal T}$,
such that $\mbox{\boldmath $\varphi$} \equiv \mbox{\boldmath
$\varphi$}(\mbox{\boldmath $\theta$},\epsilon)$ and $\|
\mbox{\boldmath $\varphi$}\|_{X} = {\rm O}(\epsilon)$ as $\epsilon
\to 0$. As a result, there exists the nonlinear vector field ${\bf
g} : {\cal T} \times \mathbb{R}^1 \mapsto \mathbb{R}^N$, such that
the Lyapunov--Schmidt bifurcation equations are
\begin{equation}
\label{g-definition} {\bf g}(\mbox{\boldmath $\theta$},\epsilon) =
{\cal P} {\bf F}(\mbox{\boldmath $\phi$}^{(0)}(\mbox{\boldmath
$\theta$})+\mbox{\boldmath $\varphi$}(\mbox{\boldmath
$\theta$},\epsilon),\epsilon) = 0.
\end{equation}
By the construction, the function ${\bf g}(\mbox{\boldmath
$\theta$},\epsilon)$ is analytic in $\epsilon \in {\cal O}(0)$ and
${\bf g}(\mbox{\boldmath $\theta$},0) = {\bf 0}$ for any
$\mbox{\boldmath $\theta$} \in {\cal T}$, such that the Taylor
series for ${\bf g}(\mbox{\boldmath $\theta$},\epsilon)$ in
$\epsilon \in {\cal O}(0)$ is given by
\begin{equation}
\label{Taylor-series-g} {\bf g}(\mbox{\boldmath $\theta$},\epsilon)
= \sum_{k=1}^{\infty} \epsilon^k {\bf g}^{(k)}(\mbox{\boldmath
$\theta$}).
\end{equation}
By the gauge symmetry, the function ${\bf g}(\mbox{\boldmath
$\theta$},\epsilon)$ satisfies the following relation:
\begin{equation}
\label{symmetry-gauge} \forall \alpha_0 \in \mathbb{R}^1, \;\;
\forall \mbox{\boldmath $\theta$} \in {\cal T} : \qquad {\bf
g}(\mbox{\boldmath $\theta$}+\alpha_0 {\bf p}_0,\epsilon) = {\bf
g}(\mbox{\boldmath $\theta$}+\alpha_0 {\bf p}_0,\epsilon),
\end{equation}
where ${\bf p}_0 = (1,1,....,1)^T \in \mathbb{R}^N$. The main
theorem of the Lyapunov--Schmidt reduction algorithm is summarized
as follows:

\begin{theorem}[{\bf Persistence}]
\label{theorem-1} The configuration $\mbox{\boldmath
$\phi$}^{(0)}(\mbox{\boldmath $\theta$})$ in (\ref{vortex-limit})
can be continued to the domain $\epsilon \in {\cal O}(0)$ if and
only if there exists a root $\mbox{\boldmath $\theta$}_* \in {\cal
T}$ of the vector field ${\bf g}(\mbox{\boldmath
$\theta$},\epsilon)$ in (\ref{g-definition}). Moreover, if the
root $\mbox{\boldmath $\theta$}_*$ is analytic in $\epsilon \in
{\cal O}(0)$ and $\mbox{\boldmath $\theta$}_* = \mbox{\boldmath
$\theta$}_0 + {\rm O}(\epsilon)$, the solution $\mbox{\boldmath
$\phi$}$ of the difference equation (\ref{3difference}) is
analytic in $\epsilon \in {\cal O}(0)$, such that
\begin{equation}
\label{Taylor-series-phi} \mbox{\boldmath $\phi$} = \mbox{\boldmath
$\phi$}^{(0)}(\mbox{\boldmath $\theta$}_*) + \mbox{\boldmath
$\varphi$}(\mbox{\boldmath $\theta$}_*,\epsilon) = \mbox{\boldmath
$\phi$}^{(0)}(\mbox{\boldmath $\theta$}_0) + \sum_{k=1}^{\infty}
\epsilon^k \mbox{\boldmath $\phi$}^{(k)}(\mbox{\boldmath
$\theta$}_0),
\end{equation}
where $\mbox{\boldmath $\phi$}^{(k)}(\mbox{\boldmath $\theta$}_0)$,
$k \in \mathbb{N}$ are independent on $\epsilon$.
\end{theorem}

Implementation of the Lyapunov--Schmidt reduction algorithm is based
on the analysis of the convergent Taylor series expansions
(\ref{Taylor-series-g}) and (\ref{Taylor-series-phi}). Let ${\cal M}
= D_{\mbox{\boldmath $\theta$}} {\bf g}(\mbox{\boldmath
$\theta$},\epsilon) : \mathbb{R}^N \mapsto \mathbb{R}^N$ be the
Jacobian matrix evaluated at the vector $\mbox{\boldmath $\theta$}
\in {\cal T}$. By the symmetry of the shift operators $\Sigma$, the
matrix ${\cal M}$ is symmetric. By the gauge transformation
(\ref{symmetry-gauge}), ${\cal M} {\bf p}_0 = {\bf 0} \in
\mathbb{R}^N$, such that the spectrum of ${\cal M}$ includes a zero
eigenvalue. If the zero eigenvalue of ${\cal M}$ is simple, zeros
$\mbox{\boldmath $\theta$}_*$ of the function ${\bf
g}(\mbox{\boldmath $\theta$},\epsilon)$ are uniquely continued in
$\epsilon$ modulo the gauge transformation (\ref{symmetry-gauge})
by using the Implicit Function Theorem for Analytic Vector Fields.

\begin{algorithm}[{\bf Persistence}]
\label{algorithm-1} Suppose that ${\bf g}^{(k)}(\mbox{\boldmath
$\theta$}) \equiv {\bf 0}$ for $k = 1,2,...,\kappa-1$ and ${\bf
g}^{(\kappa)}(\mbox{\boldmath $\theta$}) \neq {\bf 0}$ for $\kappa
\geq 1$. Let $\mbox{\boldmath $\theta$}_0$ be the root of ${\bf
g}^{(\kappa)}(\mbox{\boldmath $\theta$})$ and ${\cal M}^{(k)} =
D_{\mbox{\boldmath $\theta$}} {\bf g}^{(k)}(\mbox{\boldmath
$\theta$}_0)$ for $k \geq \kappa$.

\begin{enumerate}
\item If ${\rm Ker}({\cal M}^{(\kappa)}) = {\rm Span}({\bf p}_0)
\subset \mathbb{R}^N$, then the configuration
(\ref{vortex-limit}) is uniquely continued in $\epsilon \in {\cal
O}(0)$ modulo the gauge transformation (\ref{symmetry-gauge}).

\item Let ${\rm Ker}({\cal M}^{(\kappa)}) = {\rm Span}({\bf
p}_0,{\bf p}_1,...,{\bf p}_{d_{\kappa}}) \subset \mathbb{R}^N$
with $1 \leq d_{\kappa} \leq N-1$ and $P^{(\kappa)} : \mathbb{R}^N
\mapsto {\rm Ker}({\cal M}^{(\kappa)})$ be the projection matrix.
Then,

\begin{enumerate}
\item If ${\bf g}^{(\kappa + 1)}(\mbox{\boldmath $\theta$}_0)
\notin {\rm Range}({\cal M}^{(\kappa)})$, the configuration
(\ref{vortex-limit}) does not persist for any $\epsilon \neq 0$.

\item If ${\bf g}^{(\kappa + 1)}(\mbox{\boldmath $\theta$}_0) \in
{\rm Range}({\cal M}^{(\kappa)})$, the configuration
(\ref{vortex-limit}) is continued to the next order. Replace
\begin{eqnarray}
\nonumber {\cal M}^{(\kappa)} & \mapsto & P^{(\kappa)} {\cal
M}^{(\kappa + 1)} P^{(\kappa)}, \\ \nonumber P^{(\kappa)} &
\mapsto & P^{(\kappa + 1)} : \mathbb{R}^N \mapsto {\rm
Ker}(P^{(\kappa)} {\cal M}^{(\kappa + 1)} P^{(\kappa)}), \\
\nonumber \mbox{\boldmath $\theta$}_0 & \mapsto & \mbox{\boldmath
$\theta$}_0 - \epsilon \left( {\cal M}^{(\kappa)}\right)^{-1} {\bf
g}^{(\kappa + 1)}(\mbox{\boldmath $\theta$}_0), \\ \nonumber {\bf
g}^{(k+1)} & \mapsto & {\bf g}^{(k+2)}
\end{eqnarray}
and repeat the previous two steps.
\end{enumerate}
\end{enumerate}
If the algorithm stops at the order $K$ with $\kappa \leq K <
\infty$, conclude whether the vortex configuration continues in
$\epsilon$ or terminates at $\epsilon = 0$. If no $K < \infty$
exists, the algorithm does not converge in a finite number of
iterations.

\end{algorithm}

Few remarks regarding Algorithm \ref{algorithm-1} are in place.
For subsequent iterations of the algorithm, one needs to extend
the power series
\begin{equation}
\label{theta-expansion} \mbox{\boldmath $\theta$}_* =
\mbox{\boldmath $\theta$}_0 - \epsilon \left( {\cal
M}^{(\kappa)}\right)^{-1} {\bf g}^{(\kappa + 1)}(\mbox{\boldmath
$\theta$}_0) + {\rm O}(\epsilon^2)
\end{equation}
to higher orders of $\epsilon$. This expansion is evaluated from
the Taylor series (\ref{Taylor-series-g}). Additionally, the
situation when the algorithm does not converge in a finite number
of iterations may imply that the particular vortex configuration
persists beyond all orders of $\epsilon$ and has additional
parameters besides the parameter $\alpha_0$ in the gauge
transformation (\ref{symmetry-gauge}).

If the algorithm converges in a finite number of iterations, it
provides a binary answer on whether the configuration persists
beyond $\epsilon \neq 0$ or terminates at $\epsilon = 0$.
Simultaneously, the method enables us to predict spectral
stability of the persistent vortex configurations. To consider
stability of vortex configurations, we need to consider the
spectrum of operators ${\cal H}$ and $\sigma {\cal H}$ in the
neighborhood of the zero eigenvalue for $\epsilon \in {\cal
O}(0)$.

Using the representation (\ref{energy}) and the Taylor series
expansion (\ref{Taylor-series-phi}), we represent operator ${\cal
H}$ by its Taylor series
\begin{equation}
\label{Taylor-series-H} {\cal H} = {\cal H}^{(0)} +
\sum_{k=1}^{\infty} \epsilon^k {\cal H}^{(k)}
\end{equation}
and consider the truncated eigenvalue problem for the spectrum of
${\cal H}$:
$$
\left[ {\cal H}^{(0)} + \epsilon {\cal H}^{(1)} + ... +
\epsilon^{k-1} {\cal H}^{(k-1)} + \epsilon^k {\cal H}^{(k)} + {\rm
O}(\epsilon^{k+1}) \right] \mbox{\boldmath $\psi$} = \mu
\mbox{\boldmath $\psi$},
$$
where $\mu$ is eigenvalue and $\mbox{\boldmath $\psi$} \in X \times
X$ is an eigenvector.

Let $\mbox{\boldmath $\alpha$} \in {\rm Ker}({\cal M}^{(\kappa)})
\cap {\rm Ker}({\cal M}^{(\kappa + 1)}) \cap ... \cap {\rm
Ker}({\cal M}^{(k-1)}) \subset \mathbb{R}^N$ but $\mbox{\boldmath
$\alpha$} \notin {\rm Ker}({\cal M}^{(k)})$ for some $\kappa \leq
k \leq K$. It is clear that $\mbox{\boldmath $\alpha$}$ has
$(d_{k-1} + 1)$ arbitrary parameters, where $d_{\kappa - 1} =
N-1$. By using the projection operator ${\cal P}$ in
(\ref{projection-H-0}) and the relation (\ref{derivative-0}), we
obtain that
$$
\mbox{\boldmath $\alpha$} = {\cal P} \left( \sum_{n \in S}
\alpha_n {\bf e}_n \right), \qquad  \left( \sum_{n \in S} \alpha_n
{\bf e}_n \right) = D_{\mbox{\boldmath $\theta$}} \mbox{\boldmath
$\phi$}^{(0)}(\mbox{\boldmath $\theta$}_0) \mbox{\boldmath
$\alpha$},
$$
where $D_{\mbox{\boldmath $\theta$}}\mbox{\boldmath
$\phi$}^{(0)}(\mbox{\boldmath $\theta$}_0)$ is the Jacobian matrix
of the infinite-dimensional vector $(\mbox{\boldmath
$\phi$}^{(0)}(\mbox{\boldmath $\theta$}),\bar{\mbox{\boldmath
$\phi$}^{(0)}}(\mbox{\boldmath $\theta$}))$ with respect to
$\mbox{\boldmath $\theta$}^T$. It is clear that
$$
{\cal H}^{(0)} \mbox{\boldmath $\psi$}^{(0)} = 0, \quad
\mbox{where} \quad  \mbox{\boldmath $\psi$}^{(0)} = \sum_{n \in S}
\alpha_n {\bf e}_n = D_{\mbox{\boldmath $\theta$}} \mbox{\boldmath
$\phi$}^{(0)}(\mbox{\boldmath $\theta$}_0) \mbox{\boldmath
$\alpha$}.
$$
Moreover, the partial $(k-1)$-th sum of the power series
(\ref{Taylor-series-phi}) gives the zero of the nonlinear vector
field (\ref{nonlinear-vector-field}) up to the order ${\rm
O}(\epsilon^k)$ and has $(d_{k-1}+1)$ arbitrary parameters when
$\mbox{\boldmath $\theta$}_0$ is shifted in the direction of the
vector $\mbox{\boldmath $\alpha$}$. At the tangent space of the
nonlinear vector fields (\ref{nonlinear-vector-field}) in the
direction of $\mbox{\boldmath $\alpha$}$, the linear inhomogeneous
system
$$
{\cal H}^{(0)} \mbox{\boldmath $\psi$}^{(m)} + {\cal H}^{(1)}
\mbox{\boldmath $\psi$}^{(m-1)} + ... + {\cal H}^{(m)}
\mbox{\boldmath $\psi$}^{(0)} = {\bf 0}, \qquad m = 1,2,...,k-1
$$
has a particular solution in the form $\mbox{\boldmath
$\psi$}^{(m)} = D_{\mbox{\boldmath $\theta$}} \mbox{\boldmath
$\phi$}^{(m)}(\mbox{\boldmath $\theta$}_0) \mbox{\boldmath
$\alpha$}$ for $m = 1,2,...,k-1$. By extending the regular
perturbation series for isolated zero eigenvalues of ${\cal
H}^{(0)}$,
\begin{equation}
\label{decomposition-eigenvalue} \mbox{\boldmath $\psi$} =
\mbox{\boldmath $\psi$}^{(0)} + \epsilon \mbox{\boldmath
$\psi$}^{(1)} + ... + \epsilon^k \mbox{\boldmath $\psi$}^{(k)} +
{\rm O}(\epsilon^{k+1}), \qquad \mu = \mu_k \epsilon^k + {\rm
O}(\epsilon^{k+1}),
\end{equation}
we obtain the linear inhomogeneous equation
$$
{\cal H}^{(0)} \mbox{\boldmath $\psi$}^{(k)} + {\cal H}^{(1)}
\mbox{\boldmath $\psi$}^{(k-1)} + ... + {\cal H}^{(k)}
\mbox{\boldmath $\psi$}^{(0)} = \mu_k \mbox{\boldmath
$\psi$}^{(0)}.
$$
Applying the projection operator ${\cal P}$ and recalling the
definition (\ref{g-definition}), we find that the left-hand-side
of the linear equation reduces to the form
$$
{\cal P} \left[ {\cal H}^{(1)} D_{\mbox{\boldmath $\theta$}}
\mbox{\boldmath $\phi$}^{(k-1)}(\mbox{\boldmath $\theta$}_0) + ...
+ {\cal H}^{(k)} D_{\mbox{\boldmath $\theta$}} \mbox{\boldmath
$\phi$}^{(0)}(\mbox{\boldmath $\theta$}_0) \right] \mbox{\boldmath
$\alpha$} = D_{\mbox{\boldmath $\theta$}} {\bf
g}^{(k)}(\mbox{\boldmath $\theta$}_0) \mbox{\boldmath $\alpha$} =
{\cal M}^{(k)} \mbox{\boldmath $\alpha$},
$$
while ${\cal P} \mbox{\boldmath $\psi$}^{(0)} = \mbox{\boldmath
$\alpha$}$. Therefore, $\mu_k$ is an eigenvalue of the Jacobian
matrix ${\cal M}^{(k)}$ and $\mbox{\boldmath $\alpha$}$ is the
corresponding eigenvector. The equivalence between non-zero small
eigenvalues of ${\cal H}$ and non-zero eigenvalues of ${\cal M}$
is summarized as follows:

\begin{theorem}[{\bf Eigenvalues of ${\cal H}$}]
\label{theorem-2} Let Algorithm \ref{algorithm-1} converge at the
$K$-th order and the solution $\mbox{\boldmath $\phi$}$ in Theorem
\ref{theorem-1} persist for $\epsilon \neq 0$. Then,
\begin{eqnarray*}
\lambda_{\neq 0}(P^{(m-1)} {\cal M}^{(m)} P^{(m-1)}) \epsilon^m +
{\rm O}(\epsilon^{m + 1}) & \subset & \sigma({\cal H}), \qquad m =
\kappa,\kappa+1,...,K,
\end{eqnarray*}
where $\lambda_{\neq 0}({\cal M})$ and $\sigma({\cal H})$ denote
non-zero eigenvalues of matrix ${\cal M}$ and spectrum of operator
${\cal H}$, and $P^{(\kappa -1)}$ is an identity operator.
\end{theorem}

Similarly to the computations of small non-zero eigenvalues of
${\cal H}$, we consider eigenvalues of the spectral problem
(\ref{eigenvalue}) truncated at the $k$-th order approximation:
$$
\left[ {\cal H}^{(0)} + \epsilon {\cal H}^{(1)} + ... +
\epsilon^{k-1} {\cal H}^{(k-1)} + \epsilon^k {\cal H}^{(k)} + {\rm
O}(\epsilon^{k+1}) \right] \mbox{\boldmath $\psi$} = i \lambda
\sigma \mbox{\boldmath $\psi$}.
$$
By using relations $\hat{\bf e}_n = - i \sigma {\bf e}_n$ and
${\cal H}^{(0)} \hat{\bf e}_n = -2 \hat{\bf e}_n$, $\forall n \in
S$, we can see that the linear inhomogeneous equation
$$
{\cal H}^{(0)} \mbox{\boldmath $\varphi$}^{(0)} = 2 i \sigma
D_{\mbox{\boldmath $\theta$}}\mbox{\boldmath
$\phi$}^{(0)}(\mbox{\boldmath $\theta$}_0) \mbox{\boldmath
$\alpha$}
$$
has a solution
$$
\mbox{\boldmath $\varphi$}^{(0)} = \sum_{n \in S} \alpha_n
\hat{\bf e}_n = \mbox{\boldmath $\Phi$}^{(0)}(\mbox{\boldmath
$\theta$}_0) \mbox{\boldmath $\alpha$},
$$
where $\mbox{\boldmath $\Phi$}^{(0)}(\mbox{\boldmath $\theta$}_0)$
is the matrix extension of $\mbox{\boldmath
$\phi$}^{(0)}(\mbox{\boldmath $\theta$}_0)$, which consists of
vector columns $\hat{\bf e}_n$, $n \in S$. Similarly, there exists
a particular solution of the inhomogeneous problem
$$
{\cal H}^{(0)} \mbox{\boldmath $\varphi$}^{(m)} + {\cal H}^{(1)}
\mbox{\boldmath $\varphi$}^{(m-1)} + ... + {\cal H}^{(m)}
\mbox{\boldmath $\varphi$}^{(0)} = 2 i \sigma D^T_{\mbox{\boldmath
$\theta$}}\mbox{\boldmath $\phi$}^{(m)}(\mbox{\boldmath
$\theta$}_0) \mbox{\boldmath $\alpha$} , \qquad m = 1,2,...,k',
$$
in the form $\mbox{\boldmath $\varphi$}^{(m)} = \mbox{\boldmath
$\Phi$}^{(m)}(\mbox{\boldmath $\theta$}_0) \mbox{\boldmath
$\alpha$}$, where $k' = (k-1)/2$ if $k$ is odd and $k' = k/2-1$ if
$k$ is even. By extending the regular perturbation series for
isolated zero eigenvalue of $\sigma {\cal H}^{(0)}$,
\begin{equation}
\label{decomposition-stability} \mbox{\boldmath $\psi$} =
\mbox{\boldmath $\psi$}^{(0)} + \epsilon \mbox{\boldmath
$\psi$}^{(1)} + ... + \epsilon^{k-1} \mbox{\boldmath
$\psi$}^{(k-1)} + \frac{1}{2} \lambda \left( \mbox{\boldmath
$\varphi$}^{(0)} + \epsilon \mbox{\boldmath $\varphi$}^{(1)} + ...
+ \epsilon^{k'} \mbox{\boldmath $\varphi$}^{(k')} \right) +
\epsilon^k \mbox{\boldmath $\psi$}^{(k)} + {\rm
O}(\epsilon^{k+1}),
\end{equation}
where $\mbox{\boldmath $\psi$}^{(m)} = D_{\mbox{\boldmath
$\theta$}}\mbox{\boldmath $\phi$}^{(m)}(\mbox{\boldmath
$\theta$}_0) \mbox{\boldmath $\alpha$}$ for $m = 0,1,...,k-1$,
$\mbox{\boldmath $\varphi$}^{(m)} = \mbox{\boldmath
$\Phi$}^{(m)}(\mbox{\boldmath $\theta$}_0) \mbox{\boldmath
$\alpha$}$ for $m = 0,1,..,k'$, and $\lambda = \epsilon^{k/2}
\lambda_{k/2} + {\rm O}(\epsilon^{k/2+1})$, we obtain a linear
inhomogeneous problem for $\mbox{\boldmath $\psi$}^{(k)}$ at the
order ${\rm O}(\epsilon^k)$. When $k$ is odd, the linear problem
takes the form
\begin{equation}
\label{reduced-eigenvalue-stability-0} {\cal H}^{(0)}
\mbox{\boldmath $\psi$}^{(k)} + {\cal H}^{(1)} \mbox{\boldmath
$\psi$}^{(k-1)} + ... + {\cal H}^{(k)} \mbox{\boldmath
$\psi$}^{(0)} = \frac{i}{2} \lambda_{k/2}^2 \sigma \mbox{\boldmath
$\varphi$}^{(0)}.
\end{equation}
When $k$ is even,the linear problem takes the form
\begin{equation}
\label{reduced-eigenvalue-stability} {\cal H}^{(0)}
\mbox{\boldmath $\psi$}^{(k)} + {\cal H}^{(1)} \mbox{\boldmath
$\psi$}^{(k-1)} + ... + {\cal H}^{(k)} \mbox{\boldmath
$\psi$}^{(0)} + \frac{1}{2} \lambda_{k/2} \left( {\cal H}^{(1)}
\mbox{\boldmath $\varphi$}^{(k')} + ... + {\cal H}^{(k'+1)}
\mbox{\boldmath $\varphi$}^{(0)} \right) = \frac{i}{2}
\lambda_{k/2}^2 \sigma \mbox{\boldmath $\varphi$}^{(0)}.
\end{equation}
By using the projection operator ${\cal P}$, we establish the
equivalence between non-zero small eigenvalues of the operator
$\sigma {\cal H}$ and non-zero eigenvalues of the reduced
eigenvalue problems as follows:

\begin{theorem}[{\bf Stability}]
\label{theorem-3} Let Algorithm \ref{algorithm-1} converge at the
$K$-th order and the solution $\mbox{\boldmath $\phi$}$ in Theorem
\ref{theorem-1} persist for $\epsilon \neq 0$. Let operator ${\cal
H}$ have a small eigenvalue $\mu$ of multiplicity $d$, such that
$\mu = \epsilon^k \mu_k + {\rm O}(\epsilon^{k+1})$. Then, the
eigenvalue problem (\ref{eigenvalue}) admits $(2d)$ small
eigenvalues $\lambda$, such that $\lambda = \epsilon^{k/2}
\lambda_{k/2} + {\rm O}(\epsilon^{k/2+1})$, where non-zero values
$\lambda_{k/2}$ are found from the quadratic eigenvalue problems
\begin{eqnarray}
\label{reduced-problem-1} && \mbox{odd $k$:} \quad {\cal M}^{(k)}
\mbox{\boldmath $\alpha$} = \frac{1}{2} \lambda_{k/2}^2 \mbox{\boldmath $\alpha$}, \\
\label{reduced-problem-2} && \mbox{even $k$:} \quad {\cal M}^{(k)}
\mbox{\boldmath $\alpha$} + \frac{1}{2} \lambda_{k/2} {\cal
L}^{(k)} \mbox{\boldmath $\alpha$} = \frac{1}{2} \lambda_{k/2}^2
\mbox{\boldmath $\alpha$},
\end{eqnarray}
where ${\cal L}^{(k)} = {\cal P} \left[ {\cal H}^{(1)}
\mbox{\boldmath $\Phi$}^{(k')}(\mbox{\boldmath $\theta$}_0) + ...
+ {\cal H}^{(k'+1)} \mbox{\boldmath $\Phi$}^{(0)}(\mbox{\boldmath
$\theta$}_0) \right]$.
\end{theorem}

We note that matrix ${\cal L}^{(k)}$ must be skew-symmetric so
that all eigenvalues of the quadratic eigenvalue problem
(\ref{reduced-problem-2}) occur in pairs $\lambda_{k/2}$ and
$-\lambda_{k/2}$, which is a standard feature of linearized
Hamiltonian dynamical systems. Computations of these eigenvalues
can be achieved with a simple algorithm.

\begin{algorithm}[{\bf Stability}]
\label{algorithm-2} Suppose that the solution $\mbox{\boldmath
$\phi$}$ persists in Algorithm \ref{algorithm-1} and compute
${\cal M}^{(k)} = D_{\mbox{\boldmath $\theta$}} {\bf
g}^{(k)}(\mbox{\boldmath $\theta$}_0)$ for $\kappa \leq k \leq K$.

\begin{enumerate}
\item For each order $k$, where eigenvalues of ${\cal M}^{(k)}$
are non-zero, compute matrices ${\cal L}^{(k)}$.

\item Find roots $\lambda_{k/2}$ of the determinant equation for the quadratic eigenvalue
problems (\ref{reduced-problem-1})--(\ref{reduced-problem-2}).
\end{enumerate}
\end{algorithm}

\begin{example}
{\rm Consider vortex configurations (\ref{vortex1}) on the simple
cube (\ref{contour1}). By explicit computations, the bifurcation
equations are non-empty at the order $\kappa = 1$:
\begin{equation}
g^{(1)}_{l,j} = \sin (\theta_{l,j+1} - \theta_{l,j}) +
\sin(\theta_{l,j-1}-\theta_{l,j}) + \sin(\theta_{l+1,j} -
\theta_{l,j}), \qquad l = 0,1, \;\; j = 1,2,3,4,
\end{equation}
where $\theta_{2,j} = \theta_{0,j}$. Roots of ${\bf
g}^{(1)}(\mbox{\boldmath $\theta$})$ occur for vortex
configurations (\ref{vortex1}) with $\theta_0 = \{ 0, \pi \}$ and
$s_0 = \{ +1, -1\}$. The other vortex configurations
(\ref{vortex1}) with $\theta_0 = \{ \frac{\pi}{2}, \frac{3\pi}{2}
\}$ and $s_0 = \{ +1, -1\}$ terminate at $k = 1$ of Algorithm
\ref{algorithm-1}. The Jacobian matrix ${\cal M}^{(1)}$ has
eigenvalue $\lambda_{\neq 0}({\cal M}^{(1)}) = 2$ of multiplicity
$4$ for $\theta_0 = 0$ and $s_0 = 1$, eigenvalue $\lambda_{\neq
0}({\cal M}^{(1)}) = -2$ of multiplicity $4$ for $\theta_0 = \pi$
and $s_0 = 1$, and two eigenvalues $\lambda_{\neq 0}({\cal
M}^{(1)}) = \{ -2, 2\}$ of multiplicity $2$ for $\theta_0 = 0$ and
$s_0 = -1$. (The same result holds for $\theta_0 = \pi$ and $s_0 =
-1$ by symmetry.) Algorithm \ref{algorithm-1} does not converge at
$k = 1$, since ${\rm Ker}({\cal M}^{(1)})$ is four-dimensional in
all cases. Our results described in Example \ref{example-2}
indicate that Algorithm \ref{algorithm-1} converges at the order
$K = 6$ and the three vortex configurations above are uniquely
continued in $\epsilon \in {\rm O}(0)$ modulo the gauge
transformation (\ref{symmetry-gauge}).} \label{example-1}
\end{example}

\section{Computations of vortex configurations}
\label{section-package}

We have created the symbolic package that performs all steps of
Algorithms \ref{algorithm-1} and \ref{algorithm-2} described in
Section \ref{section-algorithm}. The user is supposed to input the
configuration of active nodes $S$ and the corresponding set of angles $\{
\theta_n \}_{n \in S}$. The package performs computations order by
order to detect if the phase configuration persists in the
Lyapunov--Schmidt reduction algorithm and if it is spectrally
stable. For all configurations (i)--(iii) described in
Section \ref{section-formalism}, we have found that Algorithm
\ref{algorithm-1} terminates at the finite order $k = K < \infty$
and the persistent configurations have $\mbox{\boldmath
$\theta$}_* = \mbox{\boldmath $\theta$}_0$ up to the order $k = K$,
where $\mbox{\boldmath $\theta$}_0$ is the root of the first
non-empty correction ${\bf g}^{(\kappa)}(\mbox{\boldmath $\theta$})$
with $\kappa \geq 1$.

The package consists of two parts. The first part performs
computations of the correction terms ${\bf
g}^{(k)}(\mbox{\boldmath $\theta$}_0)$ of the vector field ${\bf
g}(\mbox{\boldmath $\theta$},\epsilon)$ and the Jacobian matrices
${\cal M}^{(k)} = D_{\mbox{\boldmath $\theta$}} {\bf
g}^{(k)}(\mbox{\boldmath $\theta$}_0)$ for a given vector
$\mbox{\boldmath $\theta$}_0$ with $\kappa \leq k \leq K$
according to Algorithm \ref{algorithm-1}. We have confirmed that
there exists $K < \infty$ such that ${\rm dim} {\rm Ker}( {\cal
M}^{(K)} ) = 1$. We have also checked that ${\bf
g}^{(k)}(\mbox{\boldmath $\theta$}_0) \equiv 0$, $\kappa \leq k
\leq K$ for all persistent configurations, such that no extension
of the series (\ref{theta-expansion}) is necessary. All
non-persistent configurations terminate because ${\bf
g}^{(\kappa)}(\mbox{\boldmath $\theta$}_0) \neq 0$, i.e. the given
vortex configuration $\mbox{\boldmath $\theta$}_0$ is not a root
of the first non-empty correction ${\bf
g}^{(\kappa)}(\mbox{\boldmath $\theta$})$.

Non-zero eigenvalues of ${\cal M}^{(k)}$ are recorded in the first
part of the package for $\kappa \leq k \leq K$. By Theorem
\ref{theorem-2}, these non-zero eigenvalues give approximations of
the small eigenvalues of the linearized Hamiltonian ${\cal H}$ at
the order of ${\rm O}(\epsilon^k)$. The second part of the package
performs computations of eigenvalues of the quadratic eigenvalue
problems (\ref{reduced-problem-1})--(\ref{reduced-problem-2})
according to Algorithm \ref{algorithm-2}. Finding roots of the
relevant determinant equations is not computationally difficult
because the quadratic problem (\ref{reduced-problem-1}) is
diagonal in $\lambda^2$ and the quadratic problem
(\ref{reduced-problem-2}) is given typically by a matrix of small
size.

\begin{example}
{\rm Continuing Example \ref{example-1}, we have found four
non-zero eigenvalues of ${\cal M}^{(1)}$ at each persistent
configuration, which result in four pairs of small eigenvalues
$\lambda_{1/2}$ of the quadratic problem
(\ref{reduced-problem-1}). For instance, the vortex configuration
(\ref{vortex1}) with $\theta_0 = \pi$ and $s_0 = 1$ has
$\lambda_{\neq 0}({\cal M}^{(1)}) = -2$ of multiplicity $4$ and a
pair of eigenvalues $\lambda_{1/2} = \pm \sqrt{2 \lambda_{\neq
0}({\cal M}^{(1)})} = \pm 2i$ of multiplicity $4$. Continuing
Algorithms \ref{algorithm-1} and \ref{algorithm-2} to the order $k
= 2$, we find two non-zero eigenvalues of $P^{(1)} {\cal M}^{(2)}
P^{(1)}$, which result in two pairs of small eigenvalues
$\lambda_1$ of the quadratic problem (\ref{reduced-problem-2})
with a non-zero matrix ${\cal L}^{(2)}$. For the same selected
configuration, these eigenvalues are $\lambda_{\neq 0}({\cal M}_2)
= 2$ of multiplicity $2$ and a pair of eigenvalues $\lambda_{1} =
\pm 2i$ of multiplicity $2$. Continuing the algorithm, we have
found no non-zero eigenvalues for matrices $P^{(2)} {\cal
M}^{(3,4,5)} P^{(2)}$ and the last non-zero eigenvalue for the
matrix $P^{(2)} {\cal M}^{(6)} P^{(2)}$, such that $K = 6$. The
non-zero eigenvalue of $P^{(2)} {\cal M}^{(6)} P^{(2)}$ results in
a pair of small eigenvalues $\lambda_3$ of the quadratic problem
(\ref{reduced-problem-2}), where the matrix ${\cal L}^{(6)}$ was
found to be identically zero. For the same selected configuration,
the last non-zero eigenvalue is $\lambda_{\neq 0}({\cal M}^{(6)})
= -16$ and a pair of simple eigenvalues is $\lambda_3 = \pm
\sqrt{2 \lambda_{\neq 0}({\cal M}^{(6)})} = \pm 4 \sqrt{2} i$. The
two parts of the output of the symbolic computational package for
the selected vortex configuration are reproduced in Appendix A
from the outputs of the Mathematica software package. }
\label{example-2}
\end{example}

Similar to Example \ref{example-2}, we have performed computations
of all configurations (i)--(iii) listed in Section
\ref{section-formalism}. Our results are summarized in Tables 1--3
where we use the following convention: the entry $2 \times 4$ on
the first line of Table 1 denotes the non-zero eigenvalue $2$ of
algebraic multiplicity $4$ for a persistent vortex configuration.
The binary conclusions on persistence and stability of the given
vortex configuration are also listed for the reader's convenience.

\vspace{1cm}

\textbf{Table 1:}  Vortex configurations (\ref{vortex1}) on the
simple cube (\ref{contour1})

\begin{small}
\begin{tabular}{|c|c|c|c|c|c|c|c|c|}
\hline $S_1$ & Persists & \multicolumn{3}{c|}{Eigenvalues of
${\cal H}$}&\multicolumn{3}{c|}{Eigenvalues of $i \sigma {\cal
H}$}&Stable\\\cline{0-8}
&&$\epsilon$&$\epsilon^2$&$\epsilon^6$&$\epsilon^{1/2}$&$\epsilon$&$\epsilon^3$&\\
\hline\hline $\{ 0, \frac{\pi}{2}, \pi, \frac{3 \pi}{2} \}$&Yes&$\{2
\times 4 \}$&$\{2 \times 2\}$&$\{-16\}$ & $\{\pm 2 \times 4
\}$&$\{\pm 2i \times 2 \}$&$\{\pm 4i\sqrt{2}\}$&No\\
\hline $\{ 0, \frac{3\pi}{2}, \pi, \frac{\pi}{2}  \}$ &Yes&$\{-2
\times 2,2 \times 2\}$&$\{2 \times 2 \}$&$\{-16\}$&$\{\pm 2 \times
2,\pm 2i \times 2\}$&$\{\pm 2 \times 2\}$&$\{\pm 4i\sqrt{2}\}$&No\\
\hline
$\{\frac{\pi}{2}, \pi, \frac{3 \pi}{2}, 0\}$&No&&&&&&&\\
\hline
$\{\frac{\pi}{2}, 0, \frac{3\pi}{2}, \pi\}$&No&&&&&&&\\
\hline $\{\pi, \frac{3 \pi}{2},0, \frac{\pi}{2}\}$&Yes &$\{-2 \times
4 \}$&$\{2 \times 2 \}$&$\{-16\}$&$\{\pm 2i \times 4 \}$
&$\{\pm 2i \times 2\}$&$\{\pm 4i\sqrt{2}\}$&Yes\\
\hline $\{\frac{3\pi}{2},0,\frac{\pi}{2},\pi\}$&No&&&&&&& \\
\hline
\end{tabular}
\end{small}

\vspace{1cm}

\textbf{Table 2:} Vortex configurations (\ref{vortex3}) on the
double cross (\ref{contour3})

\begin{small}
\begin{tabular}{|c|c|c|c|c|c|c|}
\hline $S_1$&Persists&\multicolumn{2}{c|}{Eigenvalues of ${\cal
H}$}&\multicolumn{2}{c|}{Eigenvalues of $i \sigma {\cal
H}$}&Stable\\
\cline{0-6}&&$\epsilon^2$&$\epsilon^4$&$\epsilon$&$\epsilon^2$&\\
\hline\hline $\{ 0, \frac{\pi}{2}, \pi, \frac{3\pi}{2}\}$&Yes&$\{-2
\times 2,2 \times 2\}$&$\{-8,28 \times 2\}$ &$\{\pm 2 \times 2, \pm
2i \times 2 \}$&$\{\pm
4i,\pm 2\sqrt{14} \times 2\}$&No\\
\hline $\{0, \frac{3\pi}{2},\pi,\frac{\pi}{2}\}$&Yes&$\{-4,-2 \times
3, 2\}$&$\{-8,28\}$&$\{\pm 2,\pm 2i \times 3,\pm
2i\sqrt{2}\}$&$\{\pm 4i,\pm 2\sqrt{14}\}$&No\\
\hline
$\{\frac{\pi}{2},\pi,\frac{3\pi}{2}, 0\}$&No&&&&&\\
\hline
$\{\frac{\pi}{2}, 0, \frac{3\pi}{2},\pi\}$&No&&&&&\\
\hline $\{\pi, \frac{3 \pi}{2},0, \frac{\pi}{2}\}$&Yes&$\{-4 \times
2, -2 \times 4\}$&$\{-8\}$&$\{\pm 2i \times 4, \pm 2i\sqrt{2} \times
2\}$&$\{\pm 4i\}$&Yes\\
\hline
$\{\frac{3\pi}{2},0,\frac{\pi}{2},\pi\}$&No&&&&&\\
\hline
\end{tabular}
\end{small}

\vspace{1cm}

\textbf{Table 3:} Configurations (\ref{vortex2}) on the
diamond (\ref{contour2})

\begin{small}
\begin{tabular}{|c|c|c|c|c|c|c|c|}
\hline $S_{-1}$&$S_1$&Persists&\multicolumn{2}{c|}{Eigenvalues of
${\cal H}$}&\multicolumn{2}{c|}{Eigenvalues of $i \sigma {\cal
H}$}&Stable\\
\cline{0-7} &&&$\epsilon^2$&$\epsilon^4$&$\epsilon$&$\epsilon^2$&\\
\hline\hline
0&0&Yes&$\{-12,-6,2,2,4\}$&&$\{\pm 2 \times 2,\pm 2\sqrt{2},\pm 2i\sqrt{3},\pm 2i\sqrt{6}\}$&&No\\
\hline
0&$\frac{\pi}{2}$&No&&&&&\\
\hline 0&$\pi$&Yes&$\{-2\times 2,-5 \pm \sqrt{41}\}$&$12$&
$\{\pm 2i \times 2,\pm \sqrt{-10+2\sqrt{41}},\pm i\sqrt{10+2\sqrt{41}}\}$&$\pm 2\sqrt{6}$&No\\
\hline
$\frac{\pi}{2}$&$\frac{\pi}{2}$&No&&&&&\\
\hline
$\frac{\pi}{2}$&$\pi$&No&&&&&\\
\hline $\frac{\pi}{2}$&$\frac{3\pi}{2}$&Yes&$\{-8,-2 \times  3\}$&$-12$
&$\{\pm 2i \times 3,\pm 4i\}$&$\pm 2i\sqrt{6}$&Yes\\
\hline
\end{tabular}
\end{small}

\section{Comparison with full eigenvalue computations}
\label{section-computations}

\begin{figure}[tbp]
\begin{center}
\epsfxsize=6.5cm \epsffile{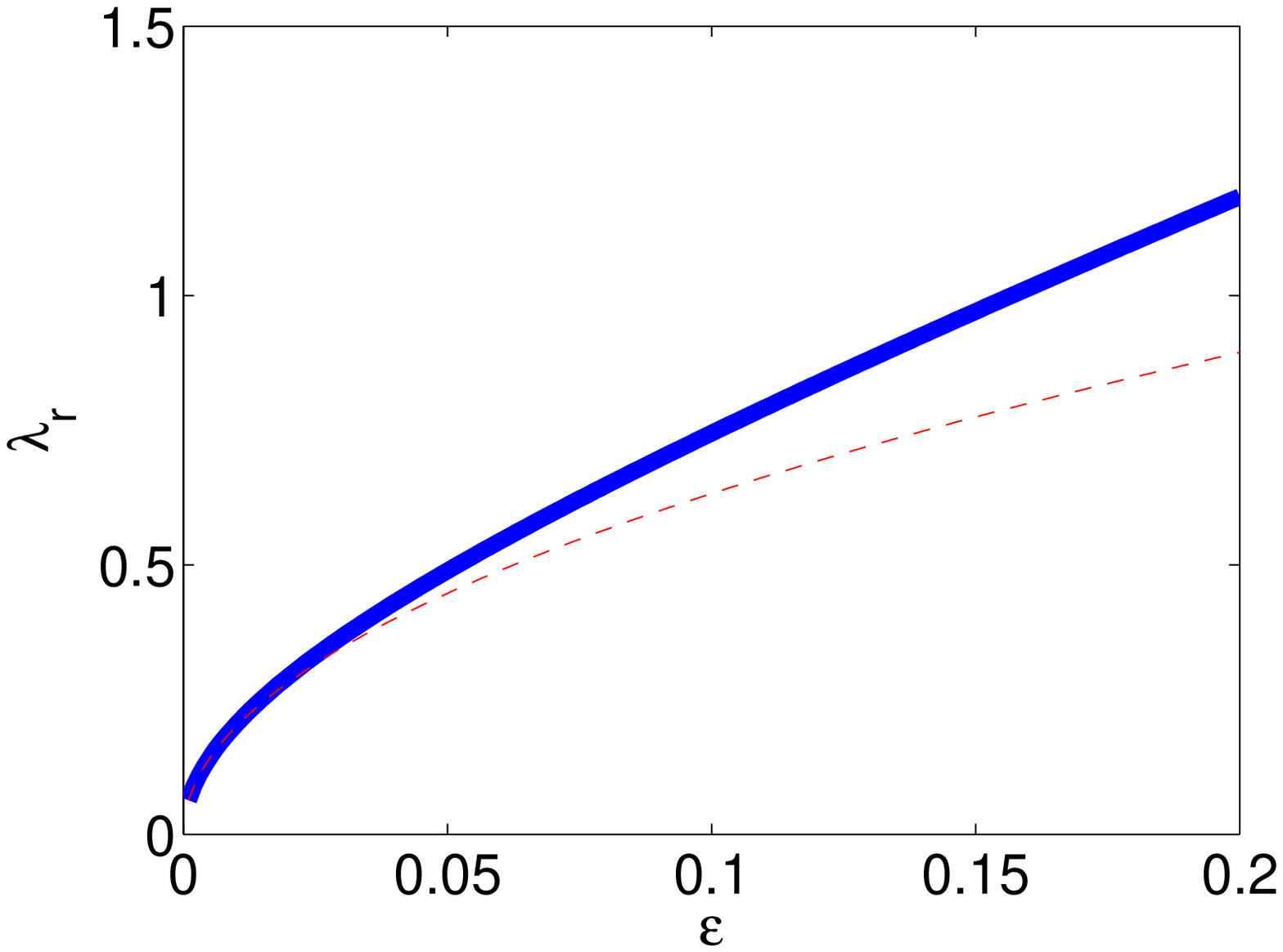}
\epsfxsize=6.5cm\epsffile{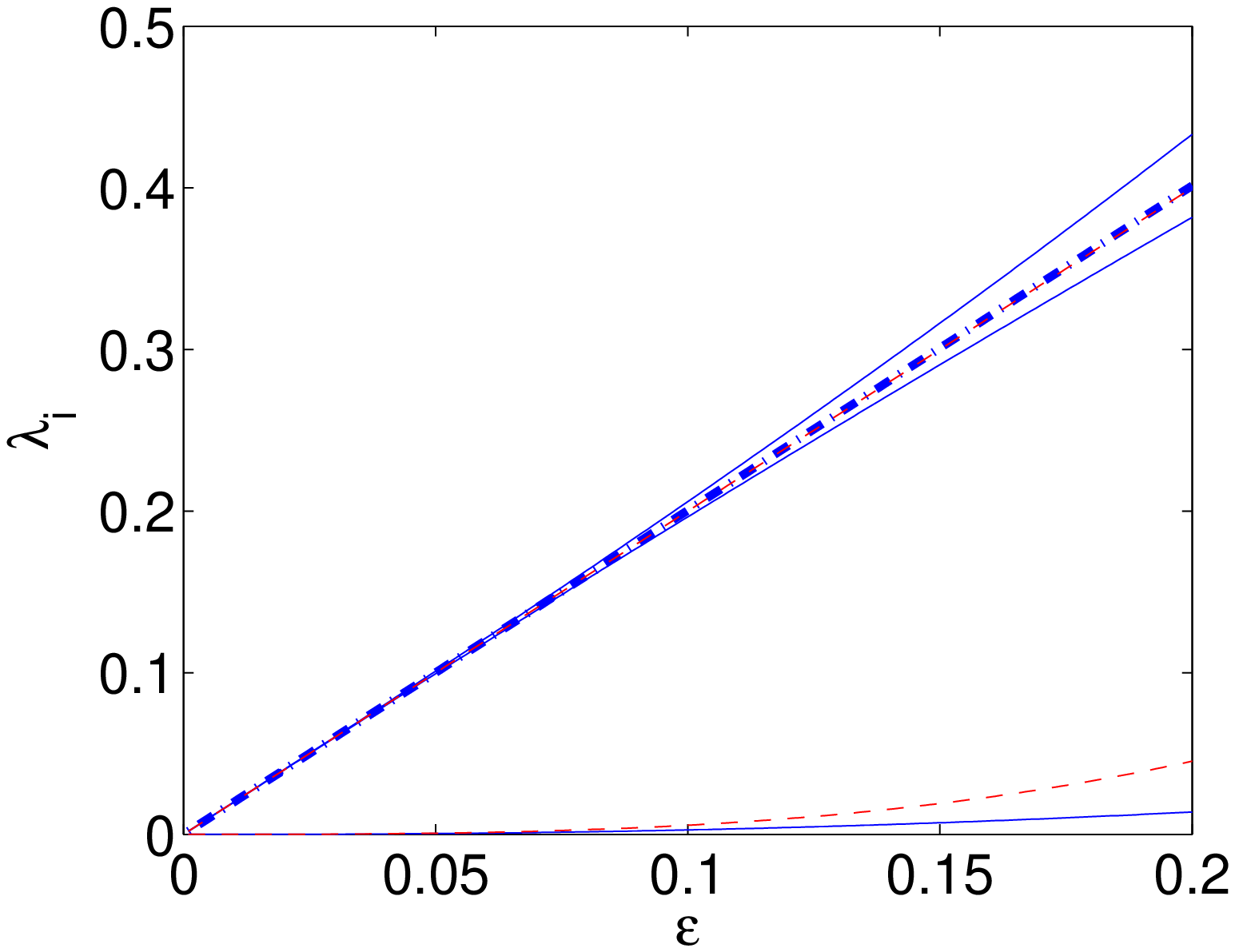} \caption{The real (left)
and imaginary (right) parts of small eigenvalues of the linearized
problem (\ref{linear-NLS}) associated with the simple cube vortex
configuration with $S_1=\{ 0, \frac{\pi}{2}, \pi, \frac{3 \pi}{2}
\}$ versus $\epsilon$; see the first paragraph of Section
\ref{section-computations} for the meaning of the different
lines.} \label{fig1}
\end{center}
\end{figure}

\begin{figure}[tbp]
\begin{center}
\epsfxsize=6.5cm \epsffile{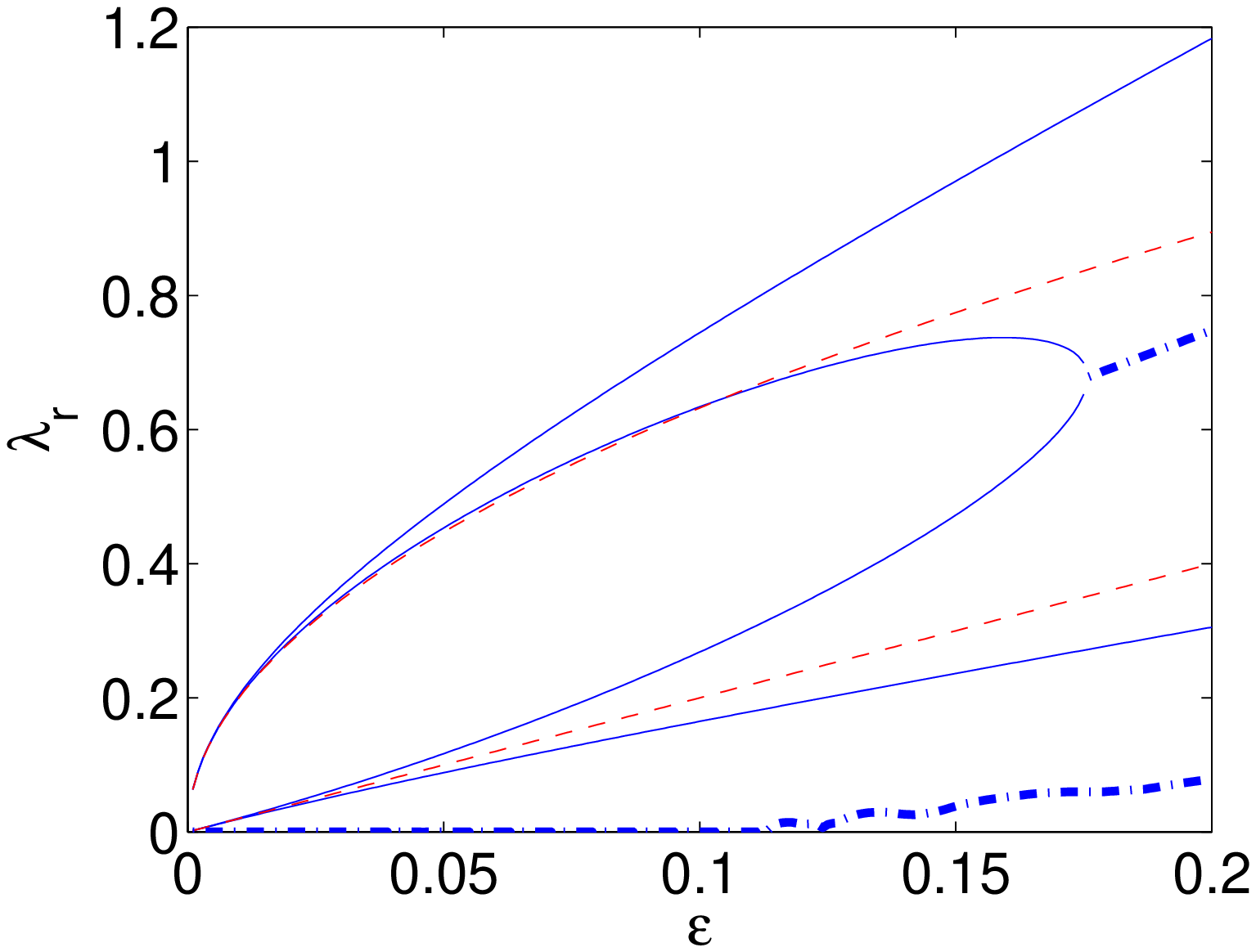}
\epsfxsize=6.5cm\epsffile{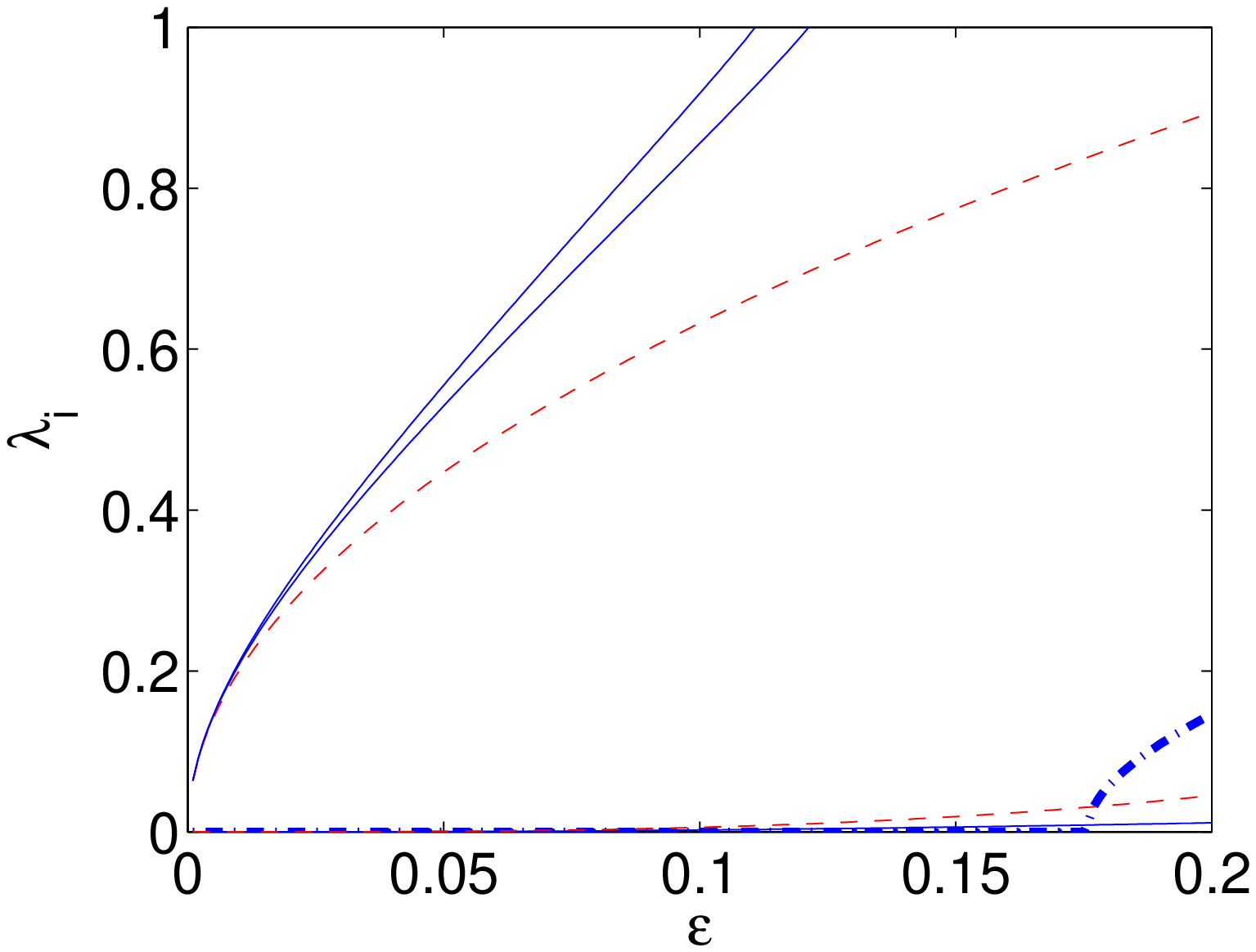} \caption{Same as in Fig.
\ref{fig1}, but for the simple cube vortex configuration with
$S_1=\{ 0, \frac{3 \pi}{2}, \pi, \frac{\pi}{2} \}$.} \label{fig2}
\end{center}
\end{figure}

\begin{figure}[tbp]
\begin{center}
\epsfxsize=6.5cm \epsffile{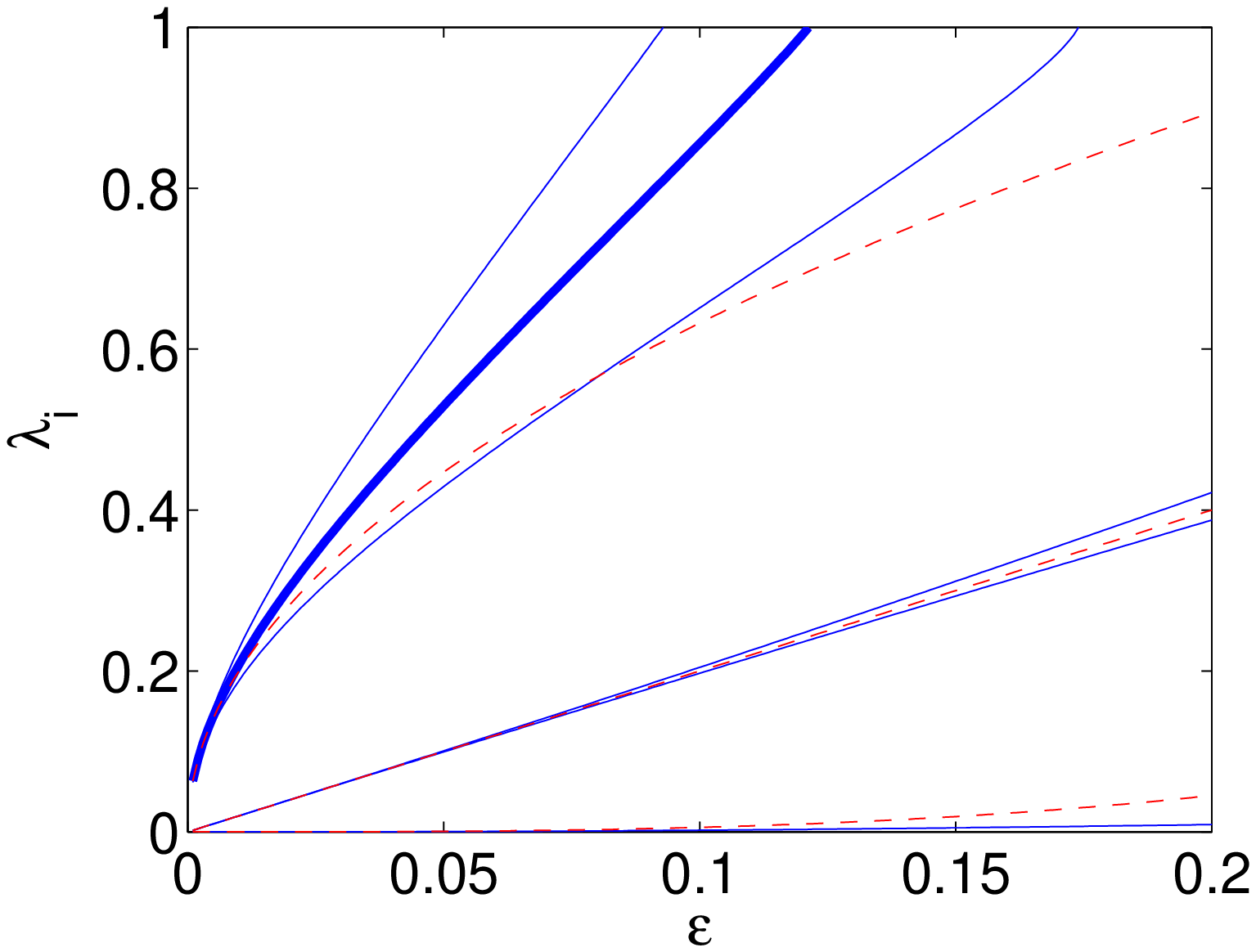} \caption{Same as in Fig.
\ref{fig1}, but for the stable simple cube vortex configuration
with $S_1=\{ \pi, \frac{3 \pi}{2}, 0, \frac{\pi}{2} \}$.}
\label{fig3}
\end{center}
\end{figure}

\begin{figure}[tbp]
\begin{center}
\epsfxsize=6.5cm \epsffile{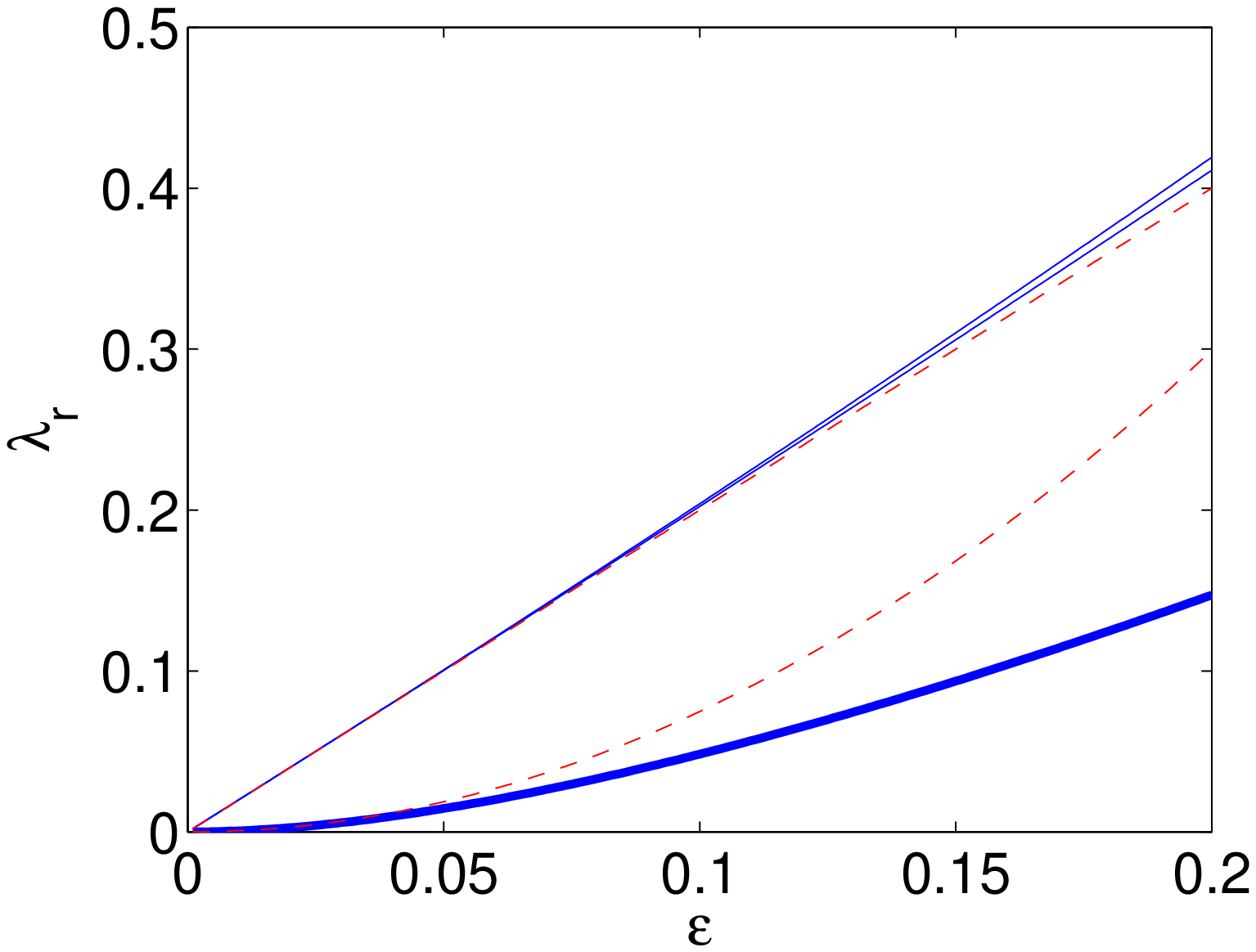}
\epsfxsize=6.5cm\epsffile{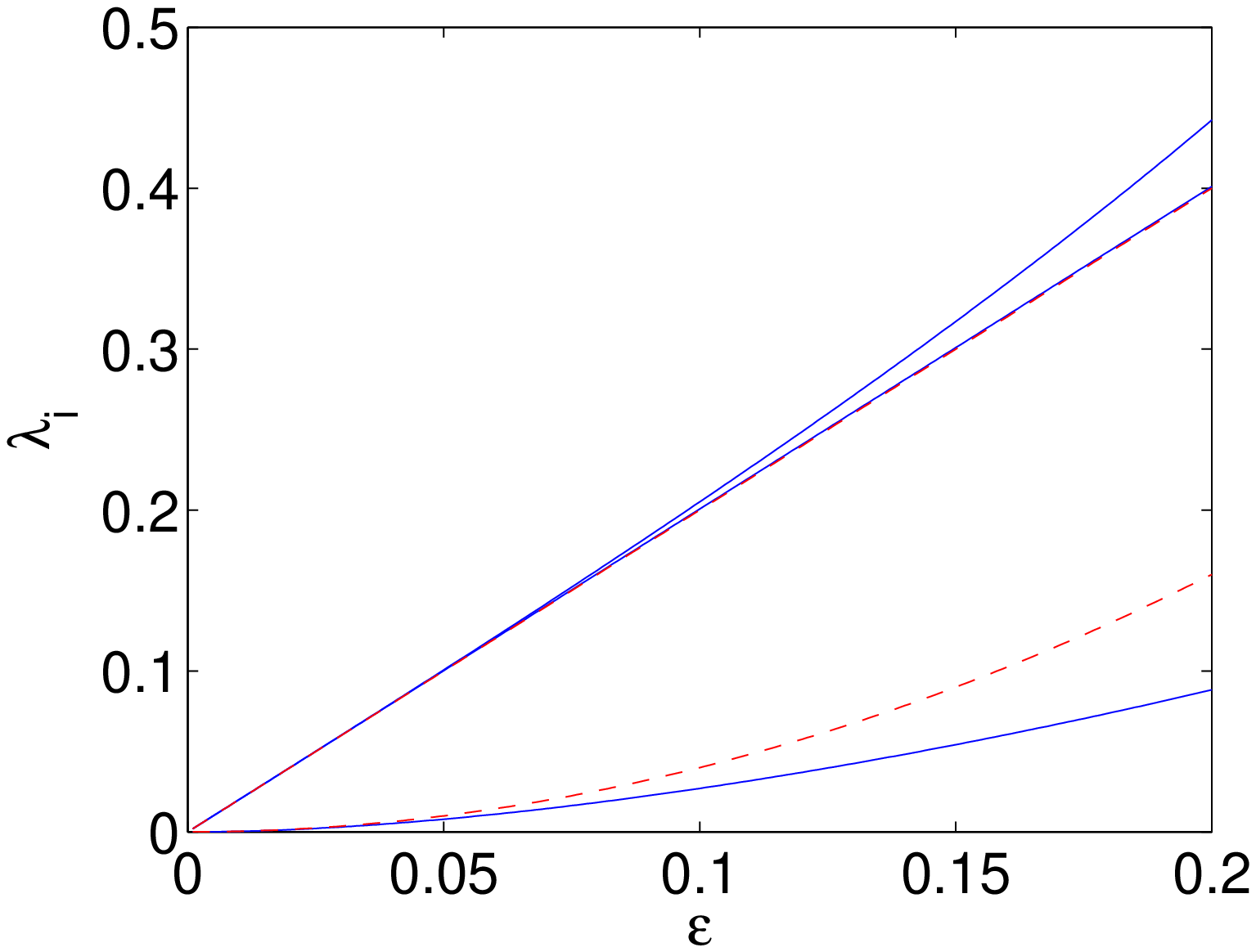} \caption{Same as Fig.
\ref{fig1}, but for the double cross vortex configuration with
$S_1=\{ 0, \frac{\pi}{2}, \pi, \frac{3 \pi}{2} \}$.} \label{fig4}
\end{center}
\end{figure}

\begin{figure}[tbp]
\begin{center}
\epsfxsize=6.5cm \epsffile{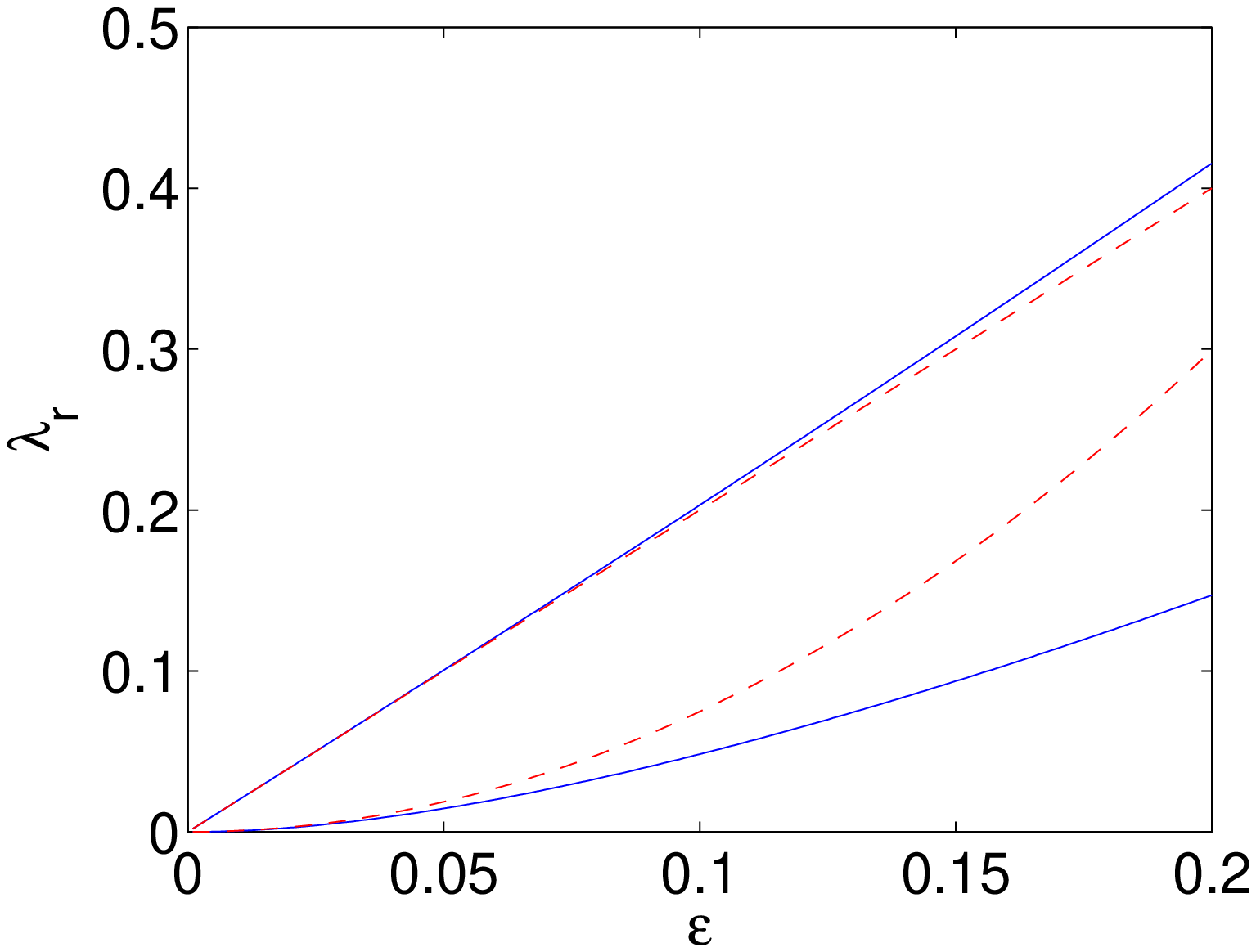}
\epsfxsize=6.5cm\epsffile{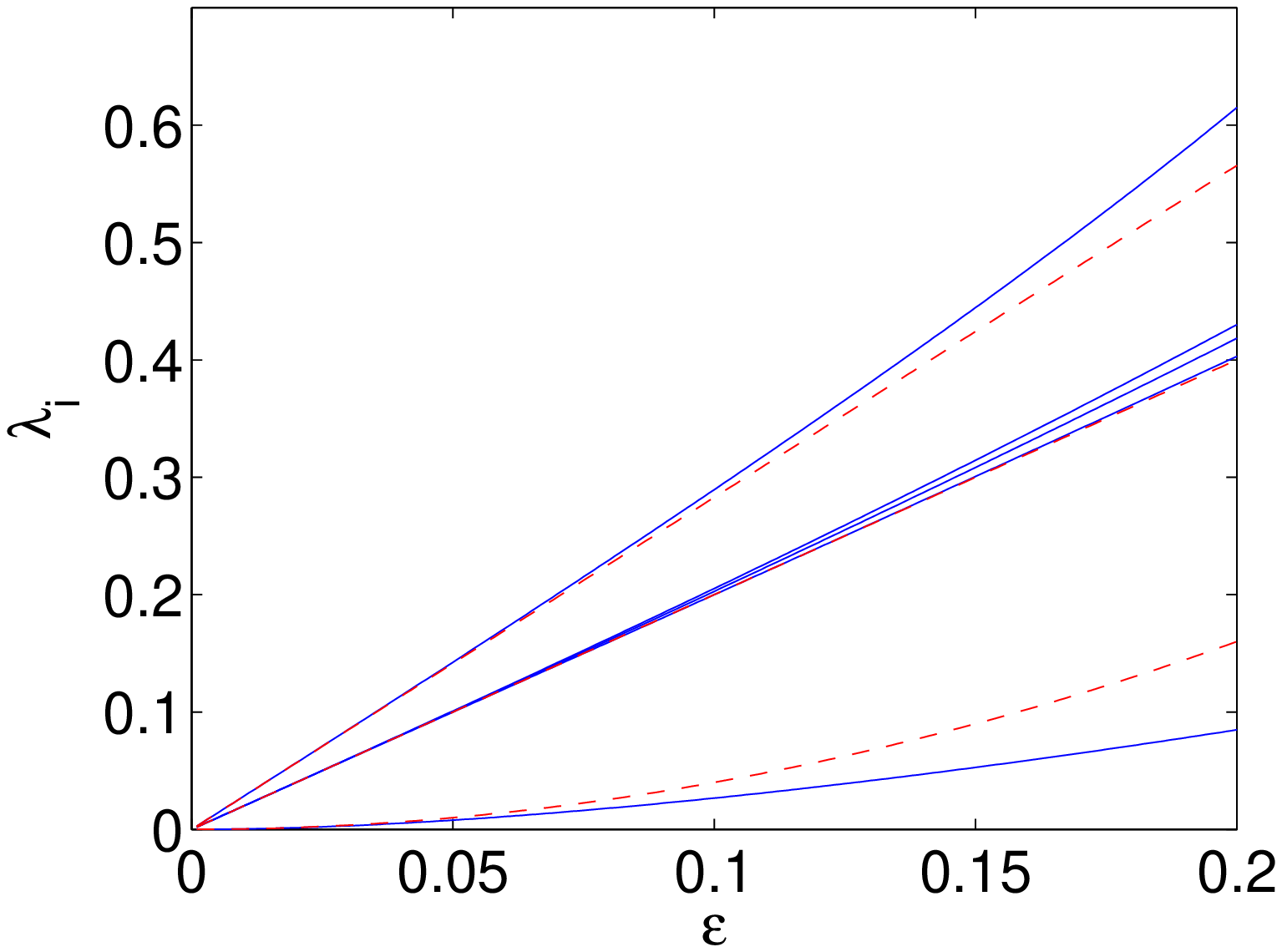} \caption{Same as Fig.
\ref{fig1}, but for the double cross vortex configuration with
$S_1=\{ 0, \frac{3 \pi}{2}, \pi, \frac{\pi}{2} \}$.} \label{fig5}
\end{center}
\end{figure}

\begin{figure}[tbp]
\begin{center}
\epsfxsize=6.5cm \epsffile{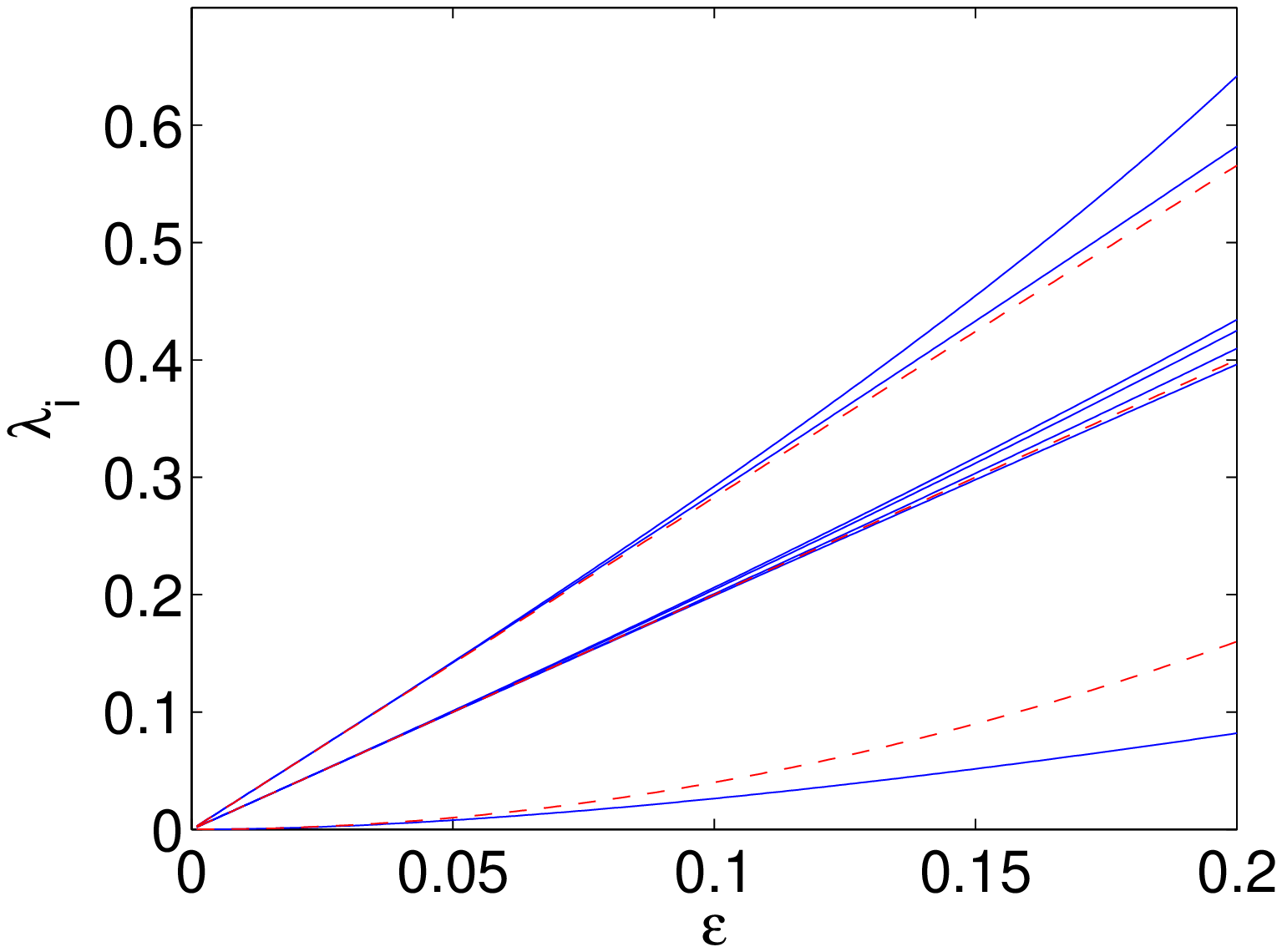} \caption{Same as Fig.
\ref{fig1}, but for the stable double cross vortex configuration
with $S_1=\{ \pi, \frac{3 \pi}{2}, 0, \frac{\pi}{2} \}$.}
\label{fig6}
\end{center}
\end{figure}

\begin{figure}[tbp]
\begin{center}
\epsfxsize=6.5cm \epsffile{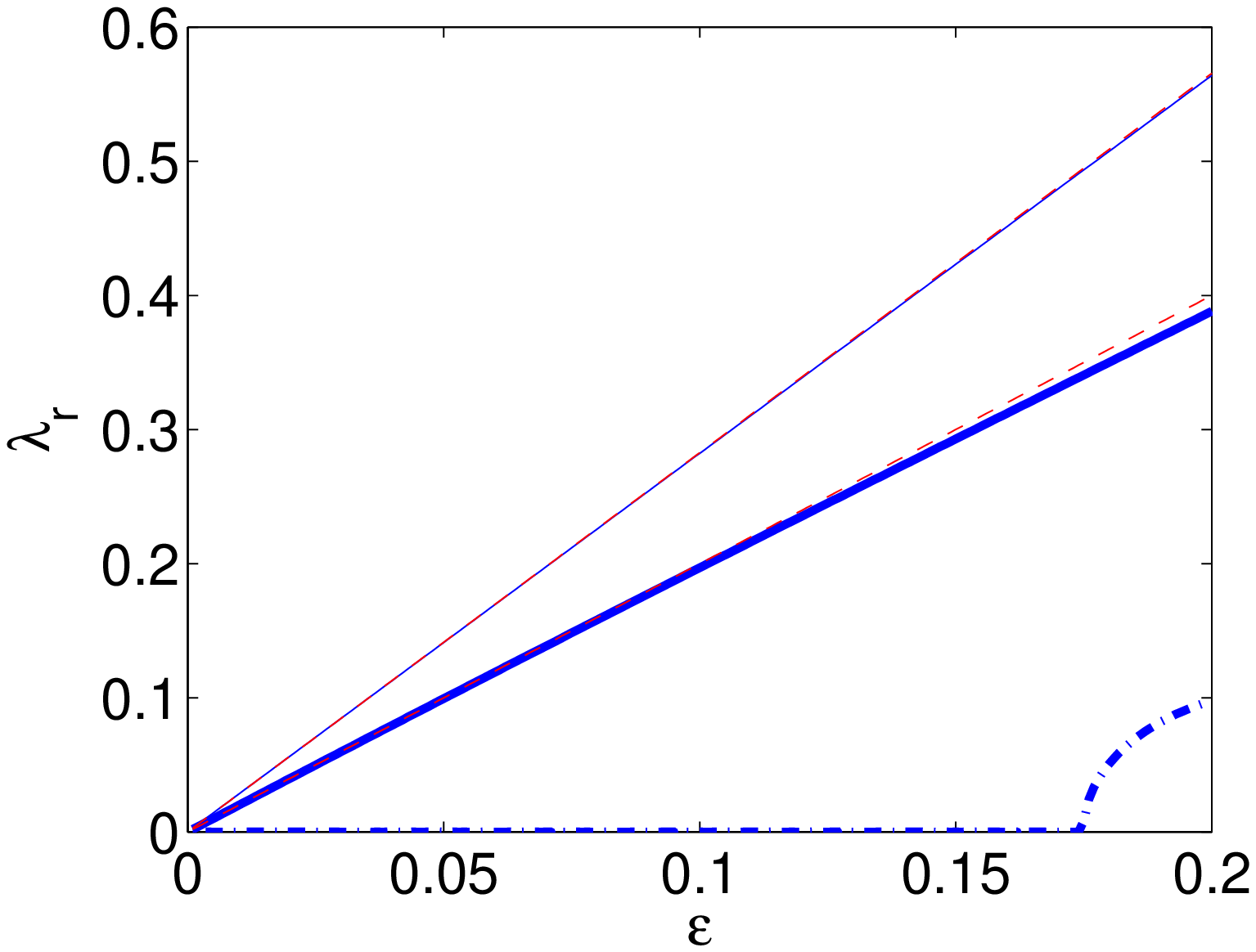}
\epsfxsize=6.5cm\epsffile{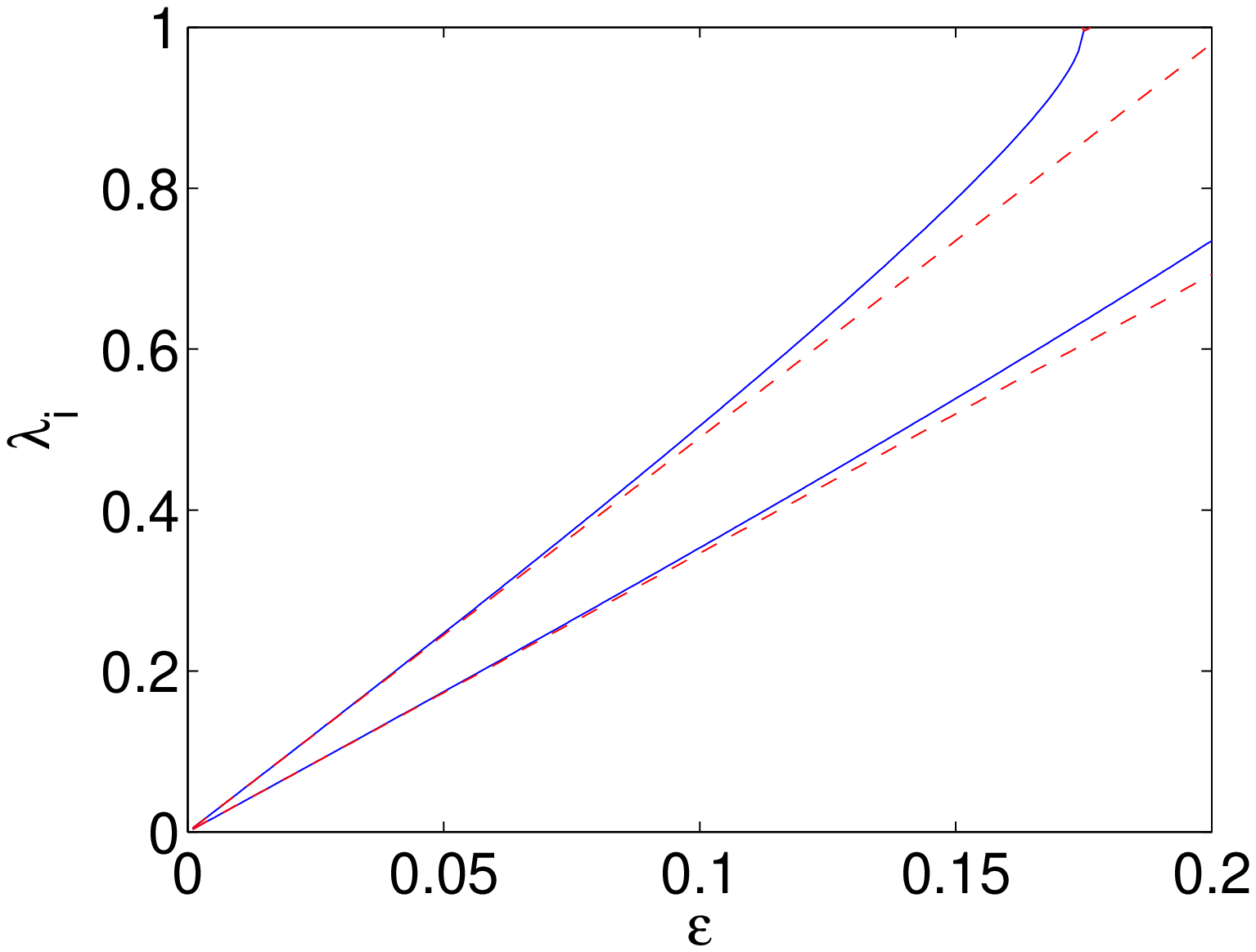} \caption{The real and
imaginary parts of the pertinent eigenvalues of the diamond
configuration with $S_{-1}=0$ and $S_1=0$ as a function of
$\epsilon$.} \label{fig7}
\end{center}
\end{figure}

\begin{figure}[tbp]
\begin{center}
\epsfxsize=6.5cm \epsffile{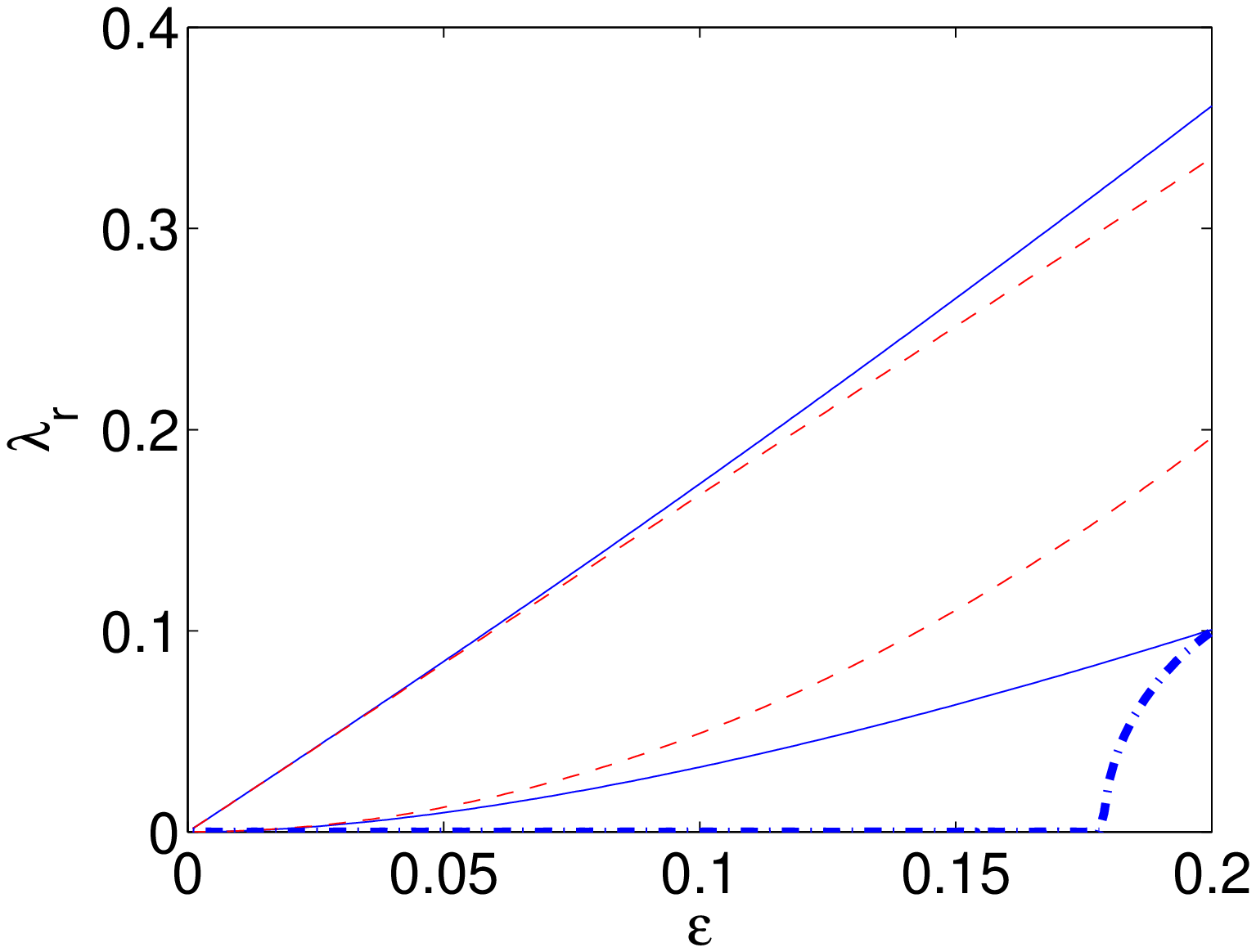}
\epsfxsize=6.5cm\epsffile{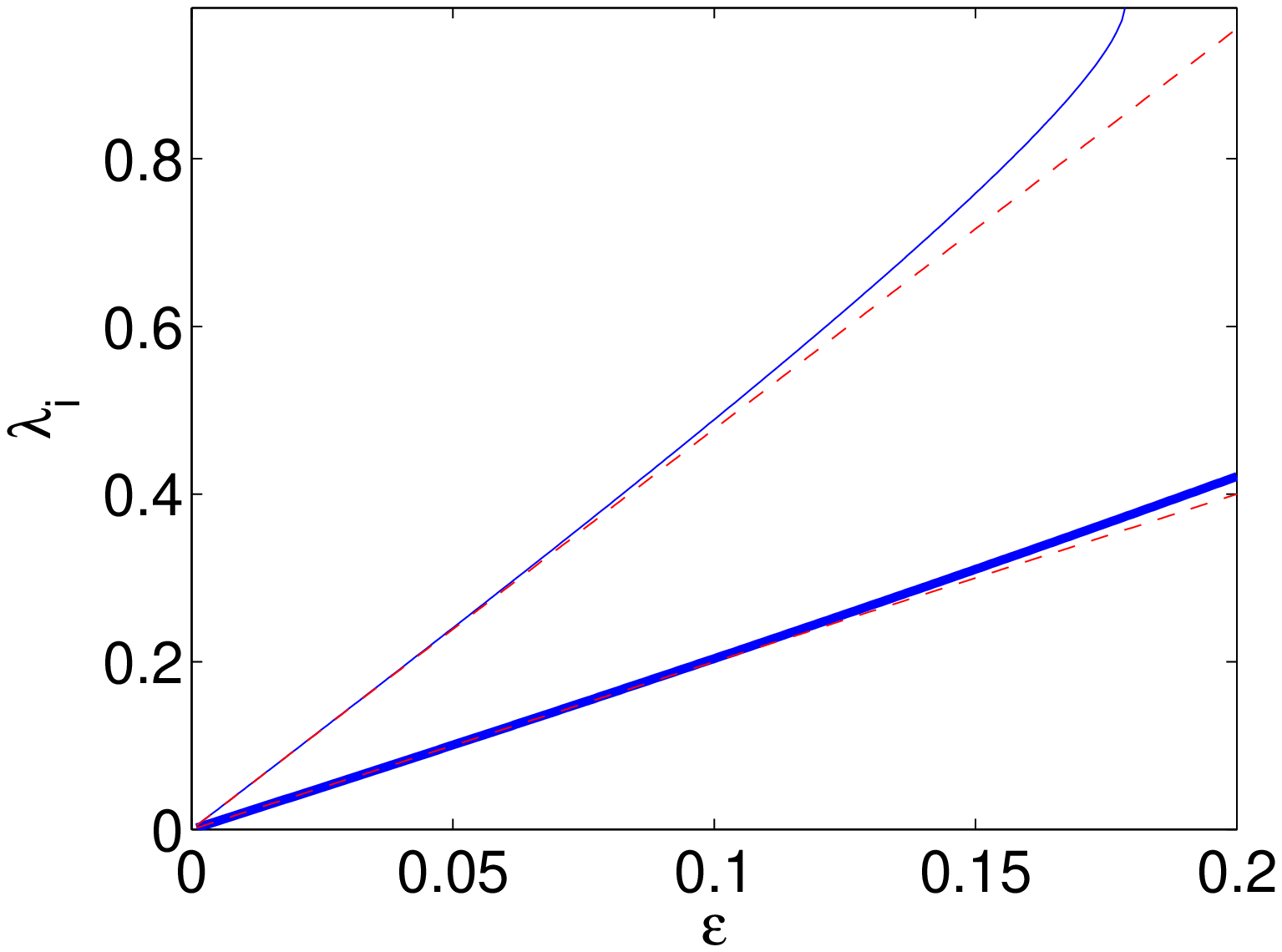} \caption{Same as in Fig.
\ref{fig7}, but for the diamond configuration with $S_{-1}=0$ and
$S_1=\pi$.} \label{fig8}
\end{center}
\end{figure}

\begin{figure}[tbp]
\begin{center}
\epsfxsize=6.5cm \epsffile{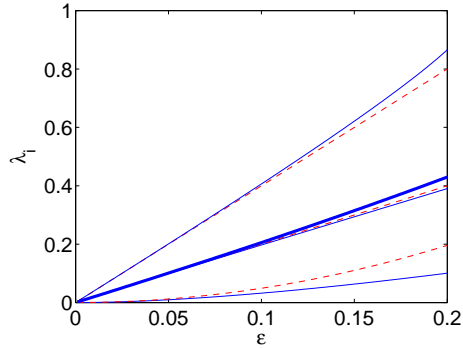} \caption{Same as in Fig.
\ref{fig7}, but for the stable diamond vortex configuration with
$S_{-1}=\frac{\pi}{2}$ and $S_1=\frac{3 \pi}{2}$.} \label{fig9}
\end{center}
\end{figure}

We proceed to test the predictions of the symbolic computations
against the results of direct numerical approximations of relevant
configurations and associated eigenvalues. Starting from the
anti-continuum limit where the solutions (\ref{vortex-limit}) are
explicit, we use numerical continuation techniques to obtain the
corresponding solutions of the difference equations
(\ref{3difference}) for finite coupling strengths of $\epsilon <
0.2$. Numerical approximations of the vortex solutions are
obtained with the fixed point iterations using Newton's method.
Once the solution $\{ \phi_n \}_{n \in \mathbb{Z}^3}$ is obtained
to the desired numerical accuracy (typically $10^{-8}$) on a
truncated numerical domain, the eigenvalue problem
(\ref{linear-NLS}) becomes a large matrix eigenvalue problem,
which is fully solved using standard numerical algebra tools. The
relevant eigenvalues with small real and imaginary parts are
isolated and their dependence on $\epsilon$ is accordingly
extracted and compared to the theoretical predictions of Tables
1--3. The results are shown in Figures \ref{fig1}-\ref{fig9},
which are discussed in more detail below. In general, we denote
numerically computed eigenvalues by solid (blue) lines, while
their counterparts from symbolic computations are plotted by
dashed (red) lines. Pairs of multiple real or imaginary
eigenvalues are denoted by thick solid lines, while quartets of
complex eigenvalues are denoted by thick dash-dotted lines.

Figure \ref{fig1} corresponds to the simple cube configuration
with $S_1=\{ 0, \frac{\pi}{2}, \pi, \frac{3 \pi}{2} \}$. As is
indicated in Table 1, this configuration should be unstable due to
an eigenvalue $\lambda \approx 2 \epsilon^{1/2}$, of multiplicity
four. What we find however is that the pair of multiple real
eigenvalue splits in two identical pairs of real eigenvalues and a
quartet of complex eigenvalues. To make things even more interesting,
all four eigenvalues in the right-half-plane have the same real
part denoted by the very thick solid line in the left panel of
Fig. \ref{fig1}. The imaginary part of the quartet of complex
eigenvalues is denoted by dash-dotted line in the right panel of
the figure. If the real part of all four eigenvalues is at the
order ${\rm O}(\epsilon^{1/2})$, the imaginary part of complex
eigenvalues occurs at the order ${\rm O}(\epsilon)$ with a
numerical approximation $\lambda \approx 2 \epsilon^{1/2} \pm 2 i
\epsilon$. Besides these four eigenvalues in the right-half-plane,
Table 1 also reports existence of a pair of double imaginary
eigenvalues at the order ${\rm O}(\epsilon)$ and a pair of simple
imaginary eigenvalues at the order ${\rm O}(\epsilon^3)$. All
these eigenvalues are shown on the right panel of Fig. \ref{fig1}
by thin lines, since the pair of double imaginary eigenvalues
splits into two pairs of simple imaginary eigenvalues. We can see
that, although the results of Table 1 give only the leading-order
approximations of small eigenvalues of the linearized problem,
they represent adequately the pattern of unstable and neutrally
stable eigenvalues.

Figure \ref{fig2} describes another simple cube vortex
configuration of Table 1 with $S_1=\{ 0, \frac{3 \pi}{2}, \pi,
\frac{\pi}{2} \}$. As theoretically predicted, we find this
configuration to be immediately unstable, due to a double pair of real
eigenvalues at the order ${\rm O}(\epsilon^{1/2})$ and another
double pair of real eigenvalues of the order ${\rm O}(\epsilon)$.
Both pairs split for small values of $\epsilon$ but remain simple
pairs of real eigenvalues for sufficiently small values of
$\epsilon$. Then, a pair of the former and one of the latter
collide for $\epsilon \approx 0.175$, leading to a quartet of
complex eigenvalues. Another double pair of imaginary eigenvalues
exists at the order ${\rm O}(\epsilon^{1/2})$ and it splits into
simple pairs of imaginary eigenvalues. When these eigenvalues meet
the continuous spectrum located at $\pm i [1,1+6 \epsilon]$, the
pairs of imaginary eigenvalues generate additional quartets of
complex eigenvalues for $\epsilon > 0.113$ and $\epsilon > 0.125$.
Finally, one more pair of imaginary eigenvalues exists at the
order ${\rm O}(\epsilon^3)$ and it remains small for $0 < \epsilon
< 0.2$.

Figure \ref{fig3} describes the third simple cube vortex
configuration of Table 1 with $S_1=\{ \pi, \frac{3 \pi}{2}, 0,
\frac{\pi}{2} \}$, which is {\it linearly stable} for small
$\epsilon$. The quadruple pair of imaginary eigenvalues at the
order ${\rm O}(\epsilon^{1/2})$ splits for small $\epsilon$ into a
double pair and two simple pairs of imaginary eigenvalues. All
these pairs generate quartets of complex eigenvalues upon
collision with the continuous spectrum for $\epsilon > 0.1$,
$\epsilon > 0.125$ and $\epsilon>0.174$. Therefore, the vortex
configuration becomes unstable for sufficiently large $\epsilon$.
A double pair of imaginary eigenvalues at the order ${\rm
O}(\epsilon)$ splits for small $\epsilon$ into simple pairs of
imaginary eigenvalues. The additional pair of imaginary
eigenvalues at the order ${\rm O}(\epsilon^3)$ remains small for
$0 < \epsilon < 0.2$.

The double cross vortex configuration of Table 2 with $S_1=\{ 0,
\frac{\pi}{2}, \pi, \frac{3 \pi}{2} \}$ is shown in Figure
\ref{fig4}. It is quite interesting that both double pairs of real
and imaginary eigenvalues at the order ${\rm O}(\epsilon)$ split
for small $\epsilon$ into simple pairs of real and imaginary
eigenvalues, while the double pair of real eigenvalues at the
order ${\rm O}(\epsilon^2)$ remain double for small $\epsilon$.

The double cross vortex configuration of Table 2 with $S_1=\{ 0,
\frac{3 \pi}{2}, \pi, \frac{\pi}{2} \}$ is shown in Fig.
\ref{fig5}. All pairs of real and imaginary eigenvalues are simple
including the triple pair of imaginary eigenvalues at the order
${\rm O}(\epsilon)$ which splits for small $\epsilon$ into three
simple pairs of imaginary eigenvalues.

Finally, the double cross vortex configuration of Table 2 with
$S_1=\{ \pi, \frac{3 \pi}{2}, 0, \frac{\pi}{2} \}$ is found to be
linearly stable for $\epsilon < 0.2$ and is shown in Fig.
\ref{fig6}. The double and quadruple pairs of imaginary
eigenvalues at the order ${\rm O}(\epsilon)$ split for small
$\epsilon$ into individual simple pairs of imaginary eigenvalues.
We note that the domain of stability of this vortex cross
configuration is wider than the one for the simple cube vortex
configuration on Fig. \ref{fig3}.

Lastly, we turn to the diamond configurations of Table 3. For the
case of $S_{-1}=0$ and $S_1=0$, shown in Fig. \ref{fig7}, the
configuration is unstable due to a simple and a double pair of
real eigenvalues, both of O$(\epsilon)$, captured very accurately
by our theoretical approximation. In addition, a complex quartet
emerges because of the collision of a pair of imaginary
eigenvalues with the continuous spectrum for $\epsilon>0.175$.

The second diamond configuration with $S_{-1}=0$ and $S_1=\pi$,
shown in Fig. \ref{fig8}, is unstable due to two simple pairs of
real eigenvalues, one at the order ${\rm O}(\epsilon)$ and one at
the order ${\rm O}(\epsilon^2)$. The simple pair of imaginary
eigenvalues becomes a quartet of complex eigenvalues upon
collision with the continuous spectrum for $\epsilon>0.179$. The
double pair of imaginary eigenvalues remains double for $0 <
\epsilon < 0.2$.

Finally, the third diamond vortex configuration with
$S_{-1}=\frac{\pi}{2}$ and $S_1=\frac{3 \pi}{2}$, shown in Fig.
\ref{fig9}, is linearly stable for $0 < \epsilon < 0.2$. In this
case also, the theoretical prediction accurately captures the
simple pair of imaginary eigenvalues at the order ${\rm
O}(\epsilon)$, the triple pair of imaginary eigenvalues (splitting
into a double pair and a simple one) at the order ${\rm
O}(\epsilon)$, and the simple pair of imaginary eigenvalues at the
order ${\rm O}(\epsilon^2)$.

\section{Conclusion}

The main benefit of the present paper is that it provides a
systematic computational approach, based on a symbolic
mathematical package to unravel the existence and linear stability
of any configuration of interest in a three-dimensional lattice
that is of paramount interest in a wide range of applications. The
package implements the conditions of the Lyapunov-Schmidt
reduction method that are necessary and, in the case of
convergence, sufficient for persistence of relevant
configurations. The reductions start from the anti-continuum limit
of the lattice and extend the solution with respect to the
coupling parameter $\epsilon$. The algorithm subsequently connects
the leading-order behavior of the small eigenvalues of the
stability problem (which bifurcate from the zero eigenvalue at
$\epsilon = 0$ and are responsible for the instability of the
vortex configurations) to the Jacobian matrix of the above
conditions. In so doing, it provides a powerful predictor that we
have always found to be accurate for small values of the coupling
parameter (typically for $\epsilon < 0.1$). We thus believe that
this work provides a valuable tool that can be used to examine
various configurations that may be of interest to both theoretical
and experimental studies in optical and soft-condensed-matter
systems.

{\bf Acknowledgement.} M. L. is supported by the NSERC USRA
scholarship. D.P. is supported by the NSERC Discovery grant and
the EPSRC Research Fellowship. P.G.K. is supported by NSF through
the grants DMS-0204585, DMS-CAREER, DMS-0505663 and DMS-0619492.

\appendix

\section{Outputs of the computational algorithm}

\subsection{Output of Algorithm 1}

Jacobian matrix at the 1st order:
$$
M1 = \left[ \begin{array}{cccccccc} -1 & 0 & 0 & 0 & 1 & 0 & 0 & 0
\\ 0 & -1 & 0 & 0 & 0 & 1 & 0 & 0
\\ 0 & 0 & -1 & 0 & 0 & 0 & 1 & 0
\\ 0 & 0 & 0 & -1 & 0 & 0 & 0 & 1
\\ 1 & 0 & 0 & 0 & -1 & 0 & 0 & 0
\\ 0 & 1 & 0 & 0 & 0 & -1 & 0 & 0
\\ 0 & 0 & 1 & 0 & 0 & 0 & -1 & 0
\\ 0 & 0 & 0 & 1 & 0 & 0 & 0 & -1
\end{array} \right]
$$

Non-zero eigenvalues of $M1$:
$$
u[1] = -2, \quad u[2] = -2, \quad u[3] = -2, \quad u[4] = -2
$$

Projection of $M2$ to Null($M1$):
$$
M2 = \left[ \begin{array}{cccc} 1 & 0 & 0 & -1
\\ 0 & 1 & -1 & 0
\\ 0 & -1 & 1 & 0
\\ -1 & 0 & 0 & 1
\end{array} \right]
$$

Non-zero eigenvalues of $M2$:
$$
u[5] = 2, \quad u[6] = 2
$$

Projection of $M3$ to Null($M2$):
$$
M3 = \left[ \begin{array}{cc} 0 & 0
\\ 0 & 0
\end{array} \right]
$$

Projection of $M4$ to Null($M2$):
$$
M4 = \left[ \begin{array}{cc} 0 & 0
\\ 0 & 0
\end{array} \right]
$$

Projection of $M5$ to Null($M2$):
$$
M5 = \left[ \begin{array}{cc} 0 & 0
\\ 0 & 0
\end{array} \right]
$$

Projection of $M6$ to Null($M2$):
$$
M6 = \left[ \begin{array}{cc} -8 & 8
\\ 8 & -8
\end{array} \right]
$$

Non-zero eigenvalues of $M6$:
$$
u[7] = -16
$$

\subsection{Output of Algorithm 2}

Projection of $i \sigma H - \lambda I$ at 1st order:
\begin{small}
$$
\left[ \begin{array}{cccccccc} -1 - \frac{1}{2} \lambda^2 & 0 & 0
& 0 & 1 & 0 & 0 & 0
\\ 0 & -1 - \frac{1}{2} \lambda^2 & 0 & 0 & 0 & 1 & 0 & 0
\\ 0 & 0 & -1  - \frac{1}{2} \lambda^2& 0 & 0 & 0 & 1 & 0
\\ 0 & 0 & 0 & -1  - \frac{1}{2} \lambda^2& 0 & 0 & 0 & 1
\\ 1 & 0 & 0 & 0 & -1  - \frac{1}{2} \lambda^2& 0 & 0 & 0
\\ 0 & 1 & 0 & 0 & 0 & -1  - \frac{1}{2} \lambda^2& 0 & 0
\\ 0 & 0 & 1 & 0 & 0 & 0 & -1  - \frac{1}{2} \lambda^2& 0
\\ 0 & 0 & 0 & 1 & 0 & 0 & 0 & -1  - \frac{1}{2} \lambda^2
\end{array} \right]
$$
\end{small}

Non-zero eigenvalues of $i \sigma H - \lambda I$ at 1st order:
$$
w[1] = -2 i, \;\; w[2] = -2 i, \;\; w[3] = -2 i,\;\; w[4] = -2 i,
\;\; w[5] = 2 i, \;\; w[6] = 2 i, \;\; w[7] = 2 i, \;\; w[8] = 2 i
$$

Projection of $i \sigma H - \lambda I$ at 2nd order:
$$
\left[ \begin{array}{cccc} 1  - \frac{1}{2} \lambda^2 & -\lambda &
\lambda & -1
\\  \lambda & 1  - \frac{1}{2} \lambda^2 & -1 &  -\lambda
\\ - \lambda & -1 & 1  - \frac{1}{2} \lambda^2 &  \lambda
\\ -1 &  \lambda & - \lambda & 1  - \frac{1}{2} \lambda^2
\end{array} \right]
$$

Non-zero eigenvalues of $i \sigma H - \lambda I$ at 2nd order:
$$
w[9] = -2 i, \;\; w[10] = -2 i, \;\;  w[11] = 2 i, \;\; w[12] = 2
i
$$

Projection of $i \sigma H - \lambda I$ at 3rd order:
$$
M3 = \left[ \begin{array}{cc} - \frac{1}{2} \lambda^2 & 0
\\ 0 & - \frac{1}{2} \lambda^2
\end{array} \right]
$$

Projection of $i \sigma H - \lambda I$ at 4th order:
$$
M4 = \left[ \begin{array}{cc} - \frac{1}{2} \lambda^2 & 0
\\ 0 & - \frac{1}{2} \lambda^2
\end{array} \right]
$$

Projection of $i \sigma H - \lambda I$ at 5th order:
$$
M5 = \left[ \begin{array}{cc} - \frac{1}{2} \lambda^2 & 0
\\ 0 & - \frac{1}{2} \lambda^2
\end{array} \right]
$$

Projection of $i \sigma H - \lambda I$ at 6th order:
$$
M6 = \left[ \begin{array}{cc} -8 - \frac{1}{2} \lambda^2& 8
\\ 8 & -8 - \frac{1}{2} \lambda^2
\end{array} \right]
$$

Non-zero eigenvalues of $i \sigma H - \lambda I$ at 6th order:
$$
w[13] = -4 i \; {\rm Sqrt}[2], \;\; w[14] = 4 i \; {\rm Sqrt}[2]
$$

\newpage

\end{document}